\documentclass[aps,prb,twocolumn,showpacs,amsmath,amssymb,superscriptaddress,longbibliography]{revtex4-1}

\usepackage{amsmath,amsfonts,bm,graphicx,verbatim,mathrsfs,units}
\usepackage[colorlinks=true,citecolor=blue]{hyperref}

\newcommand{\revision}[1]{#1}

\begin{document}

\title{Electron waiting times in coherent conductors are correlated}

 \author{David Dasenbrook}
 \thanks{These authors contributed equally to the present work.}
 \affiliation{D\'epartement de Physique Th\'eorique, Universit\'e de Gen\`eve, 1211 Gen\`eve,
  Switzerland}
 \author{Patrick P. Hofer}
 \thanks{These authors contributed equally to the present work.}
 \affiliation{D\'epartement de Physique Th\'eorique, Universit\'e de Gen\`eve, 1211 Gen\`eve,
  Switzerland}
 \author{Christian Flindt}
 \affiliation{D\'epartement de Physique Th\'eorique, Universit\'e de Gen\`eve, 1211 Gen\`eve,
  Switzerland}
  \affiliation{Department of Applied Physics, Aalto University, 00076 Aalto, Finland}

\date{\today}

\begin{abstract}
  We evaluate the joint distributions of electron waiting times in coherent conductors described by scattering theory. Successive electron waiting times in a single-channel conductor are found to be correlated due to the fermionic statistics encoded in the many-body state.  Our formalism allows us also to investigate the waiting times between charge transfer events in different outgoing channels. As an application we consider a quantum point contact in a chiral setup with one or both input channels biased by either a static or a time-dependent periodic voltage described by Floquet theory. The theoretical framework developed here can be applied to a variety of scattering problems and can in a straightforward manner be extended to joint distributions of several electron waiting times.
\end{abstract}

\pacs{72.70.+m, 73.23.-b, 73.63.-b}

% 72.70.+m Noise processes and phenomena
% 73.23.-b Electronic transport in mesoscopic systems
% 73.63.-b Electronic transport in nanoscale materials and structures

\maketitle

\section{Introduction}

Investigations of current fluctuations are useful to understand the quantum transport in small electronic conductors.\cite{blanter:2000} In the mesoscopic regime, the transport is coherent and the measurement of shot noise provides information about the effective charge and the quantum statistics of the involved quasiparticles.\cite{lesovik:1989,buttiker:1990,martin:1992,kumar:1996} A more general characterization of the transport is given by the full counting statistics (FCS) of transferred charge.\cite{levitov:1993,levitov:1996,nazarov:2003,shelankov:2003} At long times, not only the noise, but all zero-frequency current correlators can be obtained from the FCS. Besides characterizing the elementary charge transfer events,\cite{vanevic:2007,vanevic:2008,abanov:2008,abanov:2009,kambly:2009,kambly:2011} it has been realized that FCS is also intimately linked to entanglement in quantum many-body systems\cite{klich:2009,song:2011,song:2012,petrescu:2014,thomas:2015} and to fluctuation theorems at the nano-scale.\cite{forster:2008,saito:2008,esposito:2009}

\begin{figure}
  \centering
  \includegraphics[width=0.98\columnwidth]{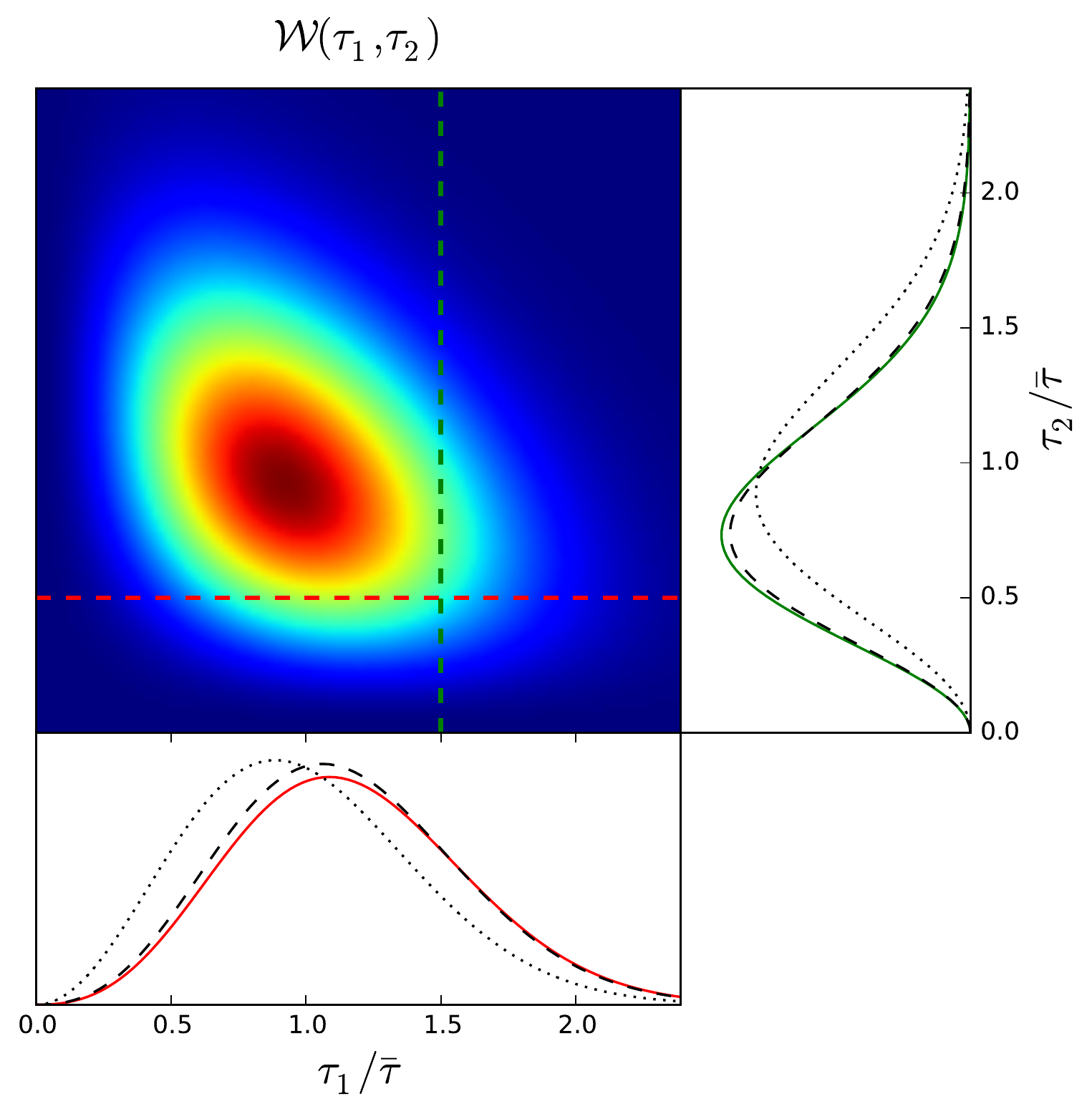}
  \caption{(Color online) Joint WTD for a fully open conduction channel. The two side panels show the joint WTD along the cuts indicated with colored dashed lines in the main panel. The joint WTD is well captured by the generalized Wigner-Dyson distribution (dashed lines) in Eq.~(\ref{eq:2WD}). The dotted lines in the side panels show the waiting time distributions if there were no correlations between subsequent waiting times. {\revision Exact results
  have been obtained using Eq.~\eqref{eq:wait2t}.}}
  \label{fig:wignerdyson}
\end{figure}

A complementary picture of the charge transport is provided by the electronic waiting time distribution (WTD). The WTD is the probability density for a waiting time of duration $\tau$ to occur between two successive charges transmitted through a conductor. In recent years, WTDs have been considered for various electronic systems, including transport through quantum dots governed by either Markovian\cite{brandes:2008,welack:2009,albert:2011,rajabi:2013,sothman:2014} or non-Markovian\cite{thomas:2013} master equations, and coherent conductors described by scattering theory\cite{albert:2012,dasenbrook:2014,haack:2014} or tight-binding models.\cite{thomas:2014} For periodically driven single-electron sources,\cite{feve:2007,dubois:2013,jullien:2014} the WTD provides a useful characterization of the regularity of the emitter and it encodes information about the shape of the emitted wave functions which is not readily accessible in the FCS.\cite{dasenbrook:2014,albert:2014} Experimentally, progress has been made towards the detection of individual electrons emitted above the Fermi sea,\cite{fletcher:2013,thalineau:2014} possibly paving the way for measurements of quantum transport on short time scales.

The interest in WTDs has so far been focused on the distribution of individual electron waiting times. Such distributions, however, do not address the question of correlations between subsequent waiting times. One may ask if the observation of one waiting time will affect the following waiting time. Some results indicate that subsequent electron waiting times indeed are correlated.\cite{albert:2012,dasenbrook:2014} However, to fully answer this question, a theory of joint WTDs is needed.

Figure~\ref{fig:wignerdyson} shows an example of a joint WTD. Using a method that we develop in this paper, we have calculated the joint distribution of waiting times $\mathcal{W}(\tau_1,\tau_2)$ between electrons transmitted through a fully open conduction channel. In Refs.~\onlinecite{albert:2012,haack:2014} it was shown that the distribution of individual electron waiting times in this case is well-approximated by the Wigner-Dyson distribution
\begin{equation}
\mathcal{W}_{\mathrm{WD}}(\tau)=  \frac{32\tau^2}{\pi^2 \bar{\tau}^3} e^{-4\tau^2/\pi\bar{\tau}^2},
\label{eq:1WD}
\end{equation}
with the mean waiting time
\begin{equation}
\bar{\tau}=\frac{h}{eV},
\label{eq:taubar}
\end{equation}
determined by the applied voltage $V$. If subsequent waiting times are uncorrelated, the joint WTD in Fig.~\ref{fig:wignerdyson} should factorize as $\mathcal{W}(\tau_1,\tau_2)=\mathcal{W}_{\mathrm{WD}}(\tau_1)\mathcal{W}_{\mathrm{WD}}(\tau_2)$. Such a factorization is often referred to as a renewal property.\cite{cox:1962} However, as we find, the joint WTD in Fig.~\ref{fig:wignerdyson} cannot be written in this simple form. This demonstrates that subsequent electron waiting times are correlated. In fact, based on the analogy between WTDs and level spacing statistics exploited in Refs.~~\onlinecite{albert:2012,haack:2014}, the joint WTD is expected to take the form\cite{herman:2007}
\begin{equation}
  \mathcal{W}_{\mathrm{WD}}(\tau_1, \tau_2) = \frac{4 b^4}{\pi \sqrt{3}\bar{\tau}^6}
  \tau_1^2 \tau_2^2 (\tau_1 + \tau_2)^2
  e^{-\frac{2b}{3\bar{\tau}^2} \left(\tau_1^2 + \tau_2^2 +
      \tau_1 \tau_2 \right)},
\label{eq:2WD}
\end{equation}
with $b=729/(128\pi)$. Figure~\ref{fig:wignerdyson} shows that our results are well-approximated by this generalized Wigner-Dyson distribution, which cannot be factorized into products of the WTD in Eq.~(\ref{eq:1WD}).

The purpose of this work is to present our method for calculating joint WTDs. The method uses a relation between WTDs and the idle time probability (ITP).\cite{albert:2012,thomas:2013,dasenbrook:2014} The ITP is the probability of detecting no electrons in a time interval of length $\tau$. For stationary processes, the WTD can be expressed as the second derivative of the ITP with respect to $\tau$. As a second-quantized formulation shows, each derivative corresponds to the detection of an electron. Building on this principle, we obtain WTDs for detections in different channels and joint distributions of successive electron waiting times.

\begin{figure}
  \centering
\includegraphics[width=.9\columnwidth]{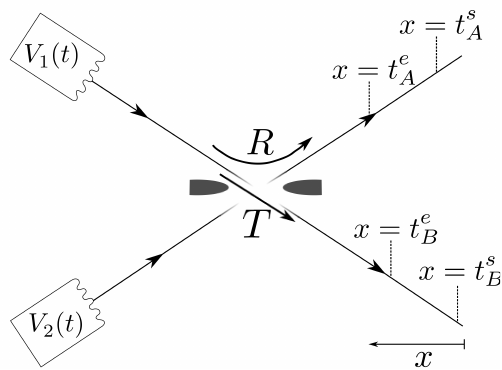}
\caption{Chiral setup with electrons in two incoming channels being partitioned on a QPC. A  (time-dependent) voltage can be applied to the contacts of the incoming channels. The QPC has transmission $T$ and reflection $R$.  The transmitted and reflected electrons are detected in the outgoing channels {\revision at different positions $x=t_{A/B}^{s/e}$ with $v_F=1$}.}
  \label{fig:setup}
\end{figure}

To illustrate our formalism, we consider a chiral setup where two incoming conduction channels are partitioned on a quantum point contact (QPC); see Fig.~\ref{fig:setup}. One or both inputs may be voltage biased, either by a static voltage or with a series of Lorentzian-shaped voltage pulses\cite{levitov:1996,ivanov:1997,keeling:2006} as recently realized experimentally.\cite{dubois:2013,jullien:2014} We calculate the joint WTD and discuss correlations between electron waiting times together with WTDs for detections in different channels. Our findings generalize earlier results for single\cite{albert:2012,dasenbrook:2014} and multi-channel\cite{haack:2014} conductors and provide insights into the correlations between subsequent electron waiting times in phase-coherent conductors. We focus here on electronic conductors, but our concepts may as well be realized in cold fermionic quantum gases.\cite{brantut:2012}

The article is organized as follows. In Sec.~\ref{sec:electronwaitingtimes} we discuss the theory of WTDs in electronic conductors and explain the concept of a renewal
process. We calculate the electronic WTD for the setup in Fig.~\ref{fig:setup}. In Sec.~\ref{sec:generalizeditp} we introduce a generalized ITP which enables us to formulate a theory of joint WTDs. In Sec.~\ref{sec:correlationsinwaitingtimes} we discuss our results for the joint WTDs with either individual electrons being partitioned on the QPC or electrons from different inputs interfering on the QPC in an electronic Hong-Ou-Mandel experiment. Finally, in Sec.~\ref{sec:conclusions} we present our conclusions. Technical details of our calculations are deferred to two appendixes at the end.

\section{Electron waiting times}
\label{sec:electronwaitingtimes}
In this section, we discuss the distribution of waiting times between successive electrons emitted into a single conduction channel. The WTD is the conditional probability density of detecting a particle at a time $t^e$ given that the last detection occurred at the earlier time $t^s$. In general, the WTD is a two-time quantity depending on both $t^s$ and~$t^e$.
% However, for systems that are translationally invariant in time, the WTD depends only on the difference of these times $\tau\equiv t^e-t^s$ and becomes a single-time quantity that we write as $\mathcal{W}(\tau)$.
{\revision We call a system stationary if the WTD depends only on the difference of
these times $\tau \equiv t^e-t^s$, and we write it as
$\mathcal{W}(\tau)$.}
We denote time intervals by $\tau$'s, and absolute times by $t$'s.

We first  define the ITP. As we will see, the ITP plays a somewhat similar role to that of a generating function: By differentiating the ITP
with respect to its time arguments, we obtain the distributions of interest, for instance the WTD. This principle becomes particularly transparent when using a second-quantized formulation of the ITP. We start by deriving the first passage time distribution from the ITP before evaluating the WTD. We illustrate our formalism by calculating the WTD for the setup depicted in Fig.~\ref{fig:setup} using either a static or a periodically modulated voltage. Finally, to gain a better understanding of correlated waiting times, we discuss the concept of a renewal process. The approach developed in this section is important as it allows us to generalize our theory to multiple conduction channels and joint WTDs later in this work.

\subsection{Idle time probability}

The ITP $\Pi(t^s, t^e)$ is the probability of observing \emph{no} transmitted electrons in the time interval $[t^s, t^e]$ at a point $x_0$ after the scatterer. To evaluate the ITP we employ second-quantization.  Working close to the Fermi level, we may linearize the dispersion relation as
\begin{equation}
\epsilon_k = \hbar v_F k.
\end{equation}
All electrons above the Fermi level thus propagate with the Fermi velocity~$v_F$ along the chiral conduction channels.  We may then evaluate the ITP by considering the operator that counts the number of transmitted particles in the spatial interval $x\in [v_F t^s, v_F t^e]$. Here $x$ is the distance to the point $x_0$ and the $x$-axis is oriented oppositely to the direction of propagation, see Fig.~\ref{fig:setup}. In second-quantization this operator reads\cite{dasenbrook:2014}
\begin{equation}
  \label{eq:Qtste}
  \widehat{Q} = \int_{v_F t^s}^{v_F t^e} \hat{b}^\dagger(x) \hat{b}(x) d x,
\end{equation}
where $\hat{b}^\dagger(x)$ and $\hat{b}(x)$ are the creation and annihilation operators of electrons at position $x$. To keep the notation simple, we have omitted the explicit time arguments of $\widehat{Q}$. We will later on make use of the relations
\begin{equation}
\begin{split}
\partial_{t^s}\widehat{Q} &= -\hat{b}^\dagger(t^s) \hat{b}(t^s),\\
\partial_{t^e}\widehat{Q} &= \hat{b}^\dagger(t^e) \hat{b}(t^e),
\end{split}
\label{eq:deriv_Q}
\end{equation}
setting $v_F=1$ throughout the rest of the paper.

The ITP can be expressed as the expectation value of the normal-ordered exponential of $-\widehat{Q}$,\cite{vyas:1988,saito:1992,levitov:1996,dasenbrook:2014}
\begin{equation}
  \label{eq:itpnormalorderedexp}
  \Pi(t^s, t^e) = \left \langle : e^{-\widehat{Q}} : \right \rangle,
\end{equation}
where the expectation value is taken with respect to the many-body state at $t=0$. {\revision This state is a non-equilibrium state consisting of a Slater determinant of single-particle scattering states in the reservoirs.\cite{hassler:2008}} To obtain a unidirectional process, we limit our analysis to particles in the transport window above the Fermi energy. Depending on the specific measurement scheme, the presence of the Fermi sea may influence the ITP. This is a subject of ongoing investigations. In this work, the notation $: \dots :$  denotes the normal-ordering of operators with respect to the Fermi sea,\citep{giuliani:2005} i.~e., all operators that create electrons above or holes below the Fermi level are moved to the left side of the operators that annihilate such excitations, with every individual permutation contributing a minus sign. We note that any time evolution of the many-body state from $t=0$ to $t=t^o$ can be absorbed into a shift of the spatial interval ($v_F=1$) from $[t^s, t^e]$ to $[t^s+t^o, t^e+t^o]$. With this in mind, we always evaluate the expectation value at $t=0$ and treat time-dependent systems by shifting the integration interval in the definition of $\widehat{Q}$ in Eq.~(\ref{eq:Qtste}).

In general, the ITP is a two-time quantity depending both on $t^s$ and $t^e$. However, for stationary processes it depends only on the time difference $\tau=t^e-t^s$ such that
\begin{equation}
  \label{eq:itpstationary}
  \Pi(t^s,t^e) = \Pi(t^e-t^s)=\Pi(\tau),
\end{equation}
just as for the WTD.

\subsection{First passage time distribution}

To appreciate the usefulness of the ITP, we first show how one can obtain the first passage time distribution $\mathcal{F}(t^s, t^e)$ from the ITP. The first passage time distribution is the probability density for the first detection of a particle to occur at the time $t^e$, given that observations were started at the earlier time $t^s$. Such a distribution has recently been evaluated for electron transport through a quantum dot, with $t^s$ being the time at which the external electronic reservoirs are connected to the quantum dot.\cite{tang:2014} To obtain the first passage time distribution from the ITP, we notice that the ITP may be expressed as
\begin{equation}
  \label{eq:itpfptd}
  \Pi(t^s,t^e) = 1 - \int_{t^s}^{t^e} \mathcal{F}(t^s, t) d t.
\end{equation}
Here, the integral equals the probability that at least one electron is observed in the interval $[t^s,t^e]$. Now, by differentiating the ITP with respect to $t^e$, we obtain
\begin{equation}
  \label{eq:fptd1}
  \mathcal{F}(t^s, t^e) = -\partial_{t^e} \Pi(t^s,t^e).
\end{equation}
Furthermore, using Eqs.~(\ref{eq:deriv_Q}) and (\ref{eq:itpnormalorderedexp}) we find
\begin{equation}
  \label{eq:fptd2}
  \mathcal{F}(t^s, t^e) = \left \langle \hat{b}^\dagger(t^e) : e^{-\widehat{Q}} : \hat{b}(t^e) \right \rangle.
\end{equation}
Equation~\eqref{eq:fptd2} is very similar to Eq.~\eqref{eq:itpnormalorderedexp}, however, the expectation value of the normal-ordered exponential is taken with respect to the many-body state with a particle removed at time $t^e$. This removal constitutes the detection event in the definition of the first passage time distribution.

A related quantity is the probability density $\overline{\mathcal{F}}(t^s,t^e)$ for detecting a particle at $t^s$ with no subsequent detections until the time $t^e$. Following the line of arguments above, we readily find
\begin{equation}
  \label{eq:Fbar1}
  \overline{\mathcal{F}}(t^s,t^e) = \partial_{t^s} \Pi(t^s,t^e)
\end{equation}
and
\begin{equation}
  \label{eq:Fbar2}
  \overline{\mathcal{F}}(t^s,t^e) = \left \langle \hat{b}^\dagger(t^s) : e^{-\widehat{Q}} : \hat{b}(t^s) \right \rangle.
\end{equation}
We see that the expectation value of the normal-ordered exponential is now taken with respect to the many-body state with a particle removed at the initial time $t^s$. For stationary problems, the first passage time only depends on the time difference $\tau\equiv t^e-t^s$ and we have
\begin{equation}
\label{eq:fptstatic}
\mathcal{F}(\tau)=\overline{\mathcal{F}}(\tau) = -\partial_\tau\Pi(\tau),
\end{equation}
as a direct consequence of Eq.~\eqref{eq:itpstationary}.

The derivations of Eqs.~(\ref{eq:fptd1},\ref{eq:fptd2},\ref{eq:Fbar1},\ref{eq:Fbar2}) illustrate an important principle that we will use in the following: By differentiating the ITP with respect to its time arguments, pairs of operators are pulled down from the exponential function corresponding to detection events at the beginning or at the end of the interval $[t^s,t^e]$.

\subsection{Waiting time distribution}

We are now in a position to derive the WTD from the ITP. We recall that the WTD is the conditional probability density of detecting a particle at a time $t^e$ given that the last detection occurred at the earlier time $t^s$. The joint probability density of detecting a particle both at $t^s$ and $t^e$ with no detection events in between is equal to the WTD multiplied by the probability density of a detection at $t^s$. The joint probability density can be obtained by differentiating the ITP with respect to both the initial time $t^s$ and the final time $t^e$. Moreover, for uni-directional charge transport, the probability density of a detection at $t^s$ is simply the average particle current $I(t^s)$ at time $t^s$. We then find
\begin{equation}
  \label{eq:wtditp1}
  I(t^s) \mathcal{W}(t^s, t^e) = -\partial_{t^e}\partial_{t^s}  \Pi(t^s,t^e),
\end{equation}
where the minus sign comes together with the partial derivative with respect to $t^e$, c.f.~Eq.~(\ref{eq:fptd1}). Using Eq.~(\ref{eq:itpnormalorderedexp}) we now arrive at the second quantized expression
\begin{equation}
  \label{eq:wtditp2}
  I(t^s) \mathcal{W}(t^s, t^e) = \left \langle \hat{b}^\dagger(t^e) \hat{b}^\dagger(t^s) : e^{-\widehat{Q}} : \hat{b}(t^s) \hat{b}(t^e) \right \rangle.
\end{equation}

For stationary processes, the average particle current equals the inverse mean waiting time, $I(t^s)=1/\langle\tau\rangle$. Combining Eqs.~(\ref{eq:itpstationary}) and (\ref{eq:wtditp1}), we then have
\begin{equation}
  \label{eq:wtdfromaverageditp}
  \mathcal{W}(\tau) = \langle \tau \rangle \partial_\tau^2 \Pi(\tau),
\end{equation}
as previously found in Refs.~\onlinecite{albert:2012,haack:2014}. For conductors driven with period $\mathcal{T}$, we define a one-time ITP by averaging over all possible starting times $t^s$,\cite{dasenbrook:2014,albert:2014} keeping the interval $\tau \equiv t^e - t^s$ fixed,
\begin{equation}
  \label{eq:itpaveraged}
  \Pi(\tau) = \frac{1}{\mathcal{T}}\int_0^\mathcal{T} \Pi(t^s,t^s +\tau) dt^s.
\end{equation}
This yields the relevant ITP if the observation starts at a random time. Employing Eqs.~(\ref{eq:fptstatic}, \ref{eq:wtdfromaverageditp}), we obtain one-time distributions which are independent of the detection time of the first electron. For the first passage time distribution, we find
\begin{equation}
  \label{eq:fptdaveraged}
  \begin{aligned}
  \mathcal{F}(\tau) &= \frac{1}{\mathcal{T}} \int_0^\mathcal{T}
  \mathcal{F}(t^s,t^s+\tau) d t^s\\&= \frac{1}{\mathcal{T}} \int_0^\mathcal{T}
    \overline{\mathcal{F}}(t^s,t^s+\tau) d t^s,
    \end{aligned}
\end{equation}
and for the WTD
\begin{equation}
  \label{eq:wtdaveraged}
  \mathcal{W}(\tau) = \frac{\langle \tau \rangle}{\mathcal{T}} \int_0^\mathcal{T} I(t^s)
  \mathcal{W}(t^s,t^s+\tau) d t^s.
\end{equation}

\begin{figure*}
  \centering
  \includegraphics[width=\textwidth]{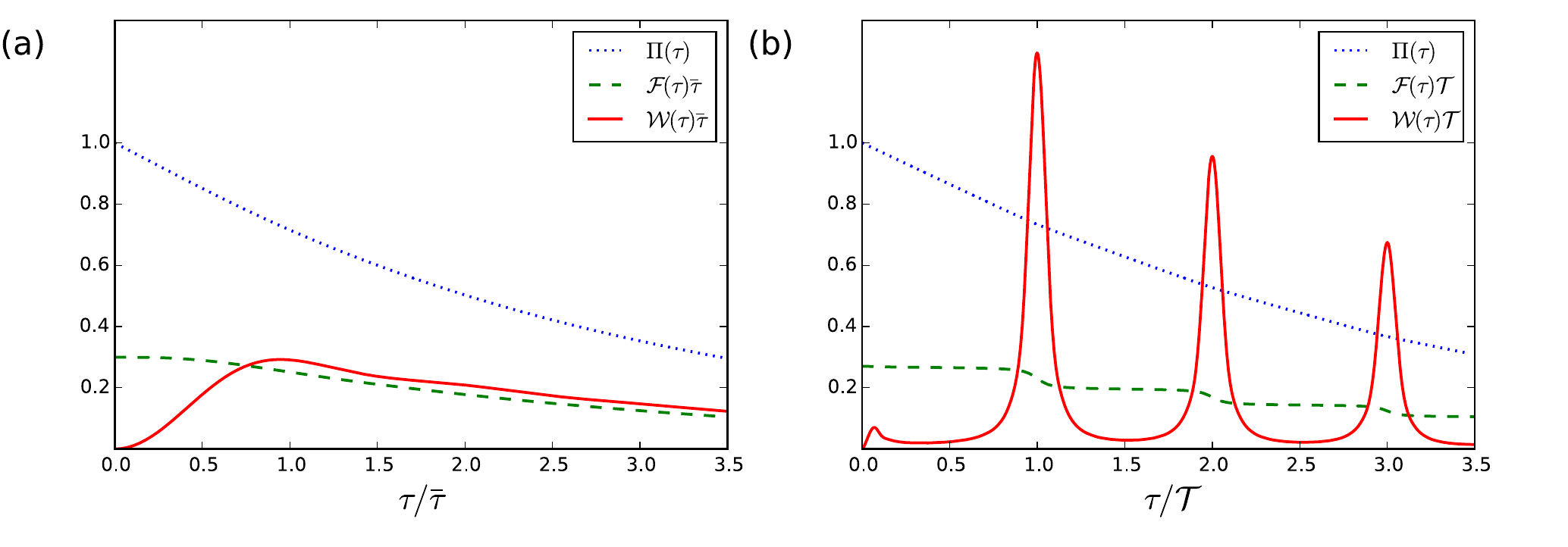}
  \caption{(Color online) Idle time probability (ITP), first passage time distribution, and waiting time distribution (WTD). (a)~Results for a static voltage applied to contact 1, $V_1(t)=V$, and no voltage applied to contact 2, $V_2(t)=0$. The QPC is tuned to $T=0.3$. We consider the electrons that are transmitted into outgoing channel $B$.
  The time is given in units of $\bar{\tau}=h/(eV)$. (b)~Results for a series of Lorentzian-shaped voltage pulses of unit charge applied to contact 1, see Eq.~\eqref{eq:levvolt}. The width of the pulses is $\Gamma=0.02\mathcal{T}$. The time is given in units of the period of the pulses~$\mathcal{T}$.}
  \label{fig:itpfptdwtd}
\end{figure*}

\subsection{Applications}
\label{sec:applications}

To illustrate our formalism, we consider the setup in Fig.~\ref{fig:setup}. Two incoming channels are partitioned on a QPC into outgoing channels labeled as $\alpha=A,B$. The electrons are non-interacting and the temperature is zero. The QPC is described by the scattering matrix
\begin{equation}
  \label{eq:hbtsmatrix}
  S = \begin{pmatrix}
    r & d \\
    -d^\ast & r^\ast
  \end{pmatrix},
\end{equation}
which relates the operators $\hat{b}_\alpha(E)$ for electrons in the outgoing channels to the operators $\hat{a}_i(E)$ for incoming electrons originating from contact
$i=1,2$. The transmission and reflection amplitudes fulfill
\begin{equation}
|d|^2+|r|^2=T+R=1
\label{eq:propcons}
\end{equation}
as a consequence of current conservation. The incoming channels are independently biased with a voltage $V_i(t)$ which is  either constant or periodic in time. For now, we keep contact $2$ grounded and we focus on the WTDs in one of the outgoing conduction channels. The generalization to multiple channels and joint WTDs follows in the next section.

The ITP in the outgoing channel $\alpha$ is given by Eq.~\eqref{eq:itpnormalorderedexp} upon attaching an index $\alpha$ to all operators
\begin{equation}
  \label{eq:itpnormalorderedexpalpha}
  \Pi_\alpha(t^s, t^e) = \left \langle : e^{-\widehat{Q}_\alpha} : \right \rangle.
\end{equation}
Using the scattering matrix, we may express the $\widehat{Q}_\alpha$'s in terms of the operators for the incoming channels. Specifically, we have
\begin{equation}
\label{eq:qops}
\widehat{Q}_A=R\int\limits_{t^s}^{t^e}\hat{a}^\dagger_1(x)\hat{a}_1(x)dx\equiv R\widehat{Q}_1,
\end{equation}
and
\begin{equation}
\widehat{Q}_B=T\widehat{Q}_1.
\label{eq:qops2}
\end{equation}
{\revision The particles in the incoming channels are in thermal equilibrium with the reservoirs.
The scattering matrix thus allows us to map the evalution of an expectation value in a
non-equilibrium state onto the evaluation of an equilibrium average.}
Note that the second contact is irrelevant since we are only concerned with the transport above the Fermi level.

The single-channel ITP can be evaluated by means of the determinant formula\cite{albert:2012},
\begin{equation}
  \label{eq:detformula}
  \Pi_\alpha(t^s,t^e) = \det(\mathbb{I} - \mathcal{Q}_\alpha),
\end{equation}
where $\mathcal{Q}_\alpha$ contains the single-particle matrix elements of the operator $\widehat{Q}_\alpha$ and $\mathbb{I}$ is the identity matrix. General expressions for a conductor with $n$ channels are given in Appendix \ref{sec:computeitp}.

We now compare the WTD for a static voltage and that of periodic voltage pulses applied to contact 1. With a constant voltage, the ITP is a function of the time difference only and we can directly employ Eq.~\eqref{eq:wtdfromaverageditp} to obtain the WTD. For the time-dependent case, we consider a series of Lorentzian-shaped voltage pulses motivated by recent experiments.\cite{dubois:2013,jullien:2014}  For pulses of unit charge, the voltage reads (see also Appendix \ref{app:levitons})
\begin{equation}
\label{eq:levvolt}
eV(t)=\sum\limits_{j=-\infty}^{\infty}\frac{2\hbar\Gamma}{\left(t-j\mathcal{T}\right)^2+\Gamma^2},
\end{equation}
where $\Gamma$ parameterizes the width of the pulses. Each pulse creates a clean single-electron excitation above the Fermi level without accompanying electron-hole pairs. Such excitations were first proposed theoretically by Levitov and co-workers\cite{levitov:1996,ivanov:1997,keeling:2006} and later named levitons following their experimental realization.\cite{dubois:2013,jullien:2014} Importantly, a time-dependent voltage has the effect of adding a phase to the single-particle states in the contact.\cite{pedersen:1998} We can treat this additional phase as a scattering phase which is picked up after the particles leave the contact but before they arrive at the scatterer,~i.~e.
\begin{equation}
\label{eq:smatvolt}
S_{V_i}(t)=e^{-i\frac{e}{\hbar}\int_{t_0}^t dt'V_i(t')}.
\end{equation}
Here $t_0$ is the time when the voltage is switched on.  In this way, the contacts can be treated as equilibrium reservoirs {\revision at the same chemical potential and all the effects due to the time-dependent driving are captured by the phase Eq.~\eqref{eq:smatvolt}}. In the energy basis, the effect of the periodic voltage is that particles can absorb or emit energy quanta of size $\hbar\Omega$, where $\Omega=2\pi/\mathcal{T}$ is the frequency of the driving. The Floquet scattering matrix\cite{moskalets:2002} encodes the quantum mechanical amplitudes for these processes and relates operators before and after the scattering phase as
\begin{equation}
  \label{eq:floquetSmatrix}
\hat{a}_i(E)=\sum_{n=-\infty}^{\infty}S_n\hat{a}'_i(E_{-n}).
\end{equation}
The operator $\hat{a}'_i(E)$ annihilates an equilibrium particle in contact $i$, while $\hat{a}_i(E)$ annihilates
a particle incident on the scatterer. We have also defined the energies
\begin{equation}
E_n = E + n \hbar \Omega.
\end{equation}
The $S_n$'s are obtained by Fourier transforming Eq.~\eqref{eq:smatvolt}. Based on Eq.~\eqref{eq:hbtsmatrix}, the $\hat{a}_i(E)$ operators can be related to the $\hat{b}_\alpha(E)$ operators and the ITP can be calculated analogously to the static case using Eq.~\eqref{eq:detformula} {\revision (see also Appendix \ref{sec:computeitp})}. Finally, to evaluate the WTD, we employ
Eqs.~(\ref{eq:wtdfromaverageditp},~\ref{eq:itpaveraged}).

Figure~\ref{fig:itpfptdwtd} shows the ITP, the first passage time distribution, and the WTD for the static voltage as well as for the train of voltage pulses. Interestingly, the two ITPs are nearly indistinguishable despite the very different voltages applied to contact 1. In both cases, the ITP decays monotonously with time from $\Pi(0)=1$ at $\tau=0$. Turning next to the first passage time distribution, some structure starts to be visible and the two voltage cases can now be distinguished. Finally, considering the WTDs, we see how they clearly differentiate between the two voltages. For the static voltage, the electronic wave packets are strongly overlapping and the WTD is rather structureless. By contrast, the application of Lorentzian-shaped voltage pulses leads to the emission of well-localized electron wave packets with a small overlap, giving raise to clear peaks at multiples of the period of the driving $\mathcal{T}$. For a fully transmitting QPC ($T=1$), the WTD essentially consists of a single peak around $\tau\simeq\mathcal{T}$ (see also Ref.~\onlinecite{dasenbrook:2014}).  For a QPC with a finite transmission, as in Fig.~\ref{fig:itpfptdwtd}, levitons may reflect back on the QPC, giving rise to the series of peaks in the WTD.  We note that the period and the width of the Lorentzian-shaped pulses, which determine the overall structure of the WTD, are easily tunable in an experiment.

\begin{figure*}
\includegraphics[width=\textwidth]{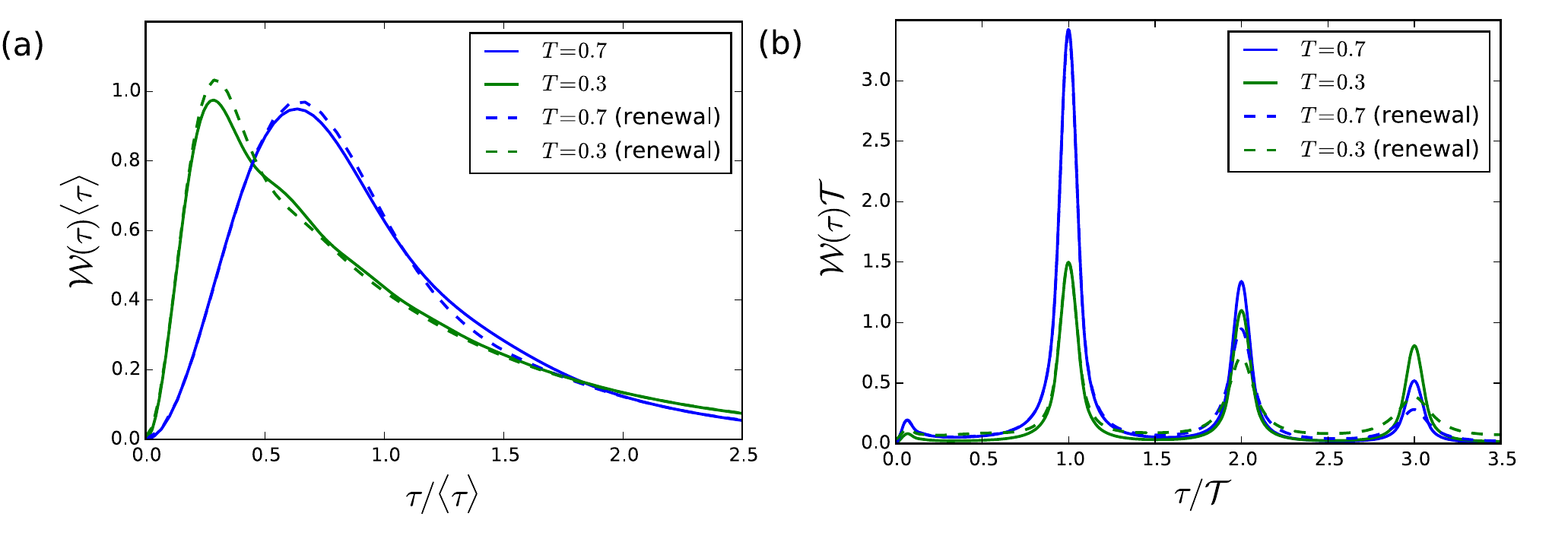}
\caption{(Color online) Comparison with renewal theory. Exact results (solid lines) for two different transmissions of the QPC are shown together with the approximation in Eq.~\eqref{eq:renewalresult} (dashed lines) based on a renewal assumption of uncorrelated waiting times.  (a)~Results for a static voltage applied to contact 1, $V_1(t)=V$, and no voltage applied to contact 2, $V_2(t)=0$. The QPC is tuned to $T=0.3$ (green lines) or $T=0.7$ (blue lines). We consider the electrons that are transmitted into outgoing channel $B$. The time is given in units of $\langle\tau\rangle=h/(TeV)$. (b)~Results for a series of Lorentzian-shaped voltage pulses of unit charge applied to contact 1. The width of the pulses is $\Gamma=0.02\mathcal{T}$. The time is given in units of the period of the pulses~$\mathcal{T}$. \label{fig:wtdfig1}}
\end{figure*}

Independently of the applied voltage, the WTD is suppressed to zero at $\tau=0$. This is a consequence of the Pauli principle which forbids two electrons to occupy the same state. For a QPC with non unit transmission $T$, the mean waiting time takes the form
\begin{equation}
\langle \tau \rangle_T=\frac{\langle \tau \rangle_{T=1}}{T},
\end{equation}
where the value at full transmission, $\langle \tau \rangle_{T=1}$, is given by $\bar{\tau}$ defined in Eq.~(\ref{eq:taubar}) for the static voltage and by the period of the driving $\mathcal{T}$ for the levitons.

\subsection{Renewal theory}
\label{sec:renewal}

It is instructive to investigate the influence of the QPC in more detail. In this connection we also introduce the concept of a renewal process. The effect of the QPC can be understood by resolving the WTD with respect to the number of reflections on the QPC that have occurred.\cite{dasenbrook:2014} Specifically, we expand the WTD as
\begin{equation}
  \label{eq:nresolvedwtd}
  \mathcal{W}(\tau) = T \sum_{n=0}^\infty R^n \mathcal{W}^{(n)}(\tau),
\end{equation}
where $\mathcal{W}^{(n)}(\tau)$ is the WTD between transmitted electrons given that $n$ reflections have occurred during the waiting time $\tau$. These $n$-resolved WTD's can be related to the joint probability distributions $\mathcal{W}_{\mathrm{in}}(\tau_1, \dots, \tau_n)$ for $n$ successive waiting times between incoming electrons, corresponding to the joint WTDs at full transmission. We have for example
\begin{equation}
\mathcal{W}^{(0)}(\tau) = \mathcal{W}_{\mathrm{in}}(\tau)
\end{equation}
and
\begin{equation}
\label{eq:w2int}
\mathcal{W}^{(1)}(\tau) =
\int_0^\tau  \mathcal{W}_{\mathrm{in}}(\tau',\tau-\tau') d\tau',
\end{equation}
where we have integrated over the time at which the reflection occurs. Similar expressions can be formulated for the $n$-resolved WTDs of higher orders in terms of the joint probability distributions for the incoming electrons.

In the following section we extend our formalism to calculations of joint WTDs. However, at this point we can proceed using a renewal assumption: We make the assumption that successive waiting times are uncorrelated and thus statistically independent and equally distributed.\cite{cox:1962} The joint probability distributions then factorize as
\begin{equation}
  \label{eq:factorizedjointwtd}
  \mathcal{W}_{\mathrm{in}}(\tau_1, \dots, \tau_n) \approx \prod_{i=1}^{n} \mathcal{W}_{\mathrm{in}}(\tau_i).
\end{equation}
Consequently, in Laplace space, we have
\begin{equation}
\mathcal{W}^{(0)}(z) = \mathcal{W}_{\mathrm{in}}(z)
\end{equation}
and
\begin{equation}
\mathcal{W}^{(1)}(z) =[\mathcal{W}_{\mathrm{in}}(z)]^2,
\end{equation}
and generally
\begin{equation}
\mathcal{W}^{(n)}(z) =[\mathcal{W}_{\mathrm{in}}(z)]^{n+1},
\end{equation}
making use of the fact the $n$-resolved WTDs under the renewal assumption are convolutions in the time domain.

Laplace transforming Eq.~\eqref{eq:nresolvedwtd}, we now obtain a geometric series which can be resummed as\cite{dasenbrook:2014}
\begin{equation}
\begin{split}
  \mathcal{W}(z) &= T \sum_{n=0}^\infty R^n [\mathcal{W}_{\mathrm{in}}(z)]^{n+1}\\
  &=\frac{T\mathcal{W}_{\mathrm{in}}(z)}{1-R\mathcal{W}_{\mathrm{in}}(z)}.
\end{split}
\label{eq:renewalresult}
\end{equation}
Returning to the time domain using an inverse Laplace transform, this expression provides us with a direct test of the renewal assumption of uncorrelated waiting times. We recall that $\mathcal{W}_{\mathrm{in}}(\tau)$ is the WTD at full transmission.

Figure~\ref{fig:wtdfig1} shows a comparison between exact results for the WTD and the approximation based on renewal theory.  Both for a static voltage and for levitons generated by periodic voltage pulses, the renewal assumption captures the main qualitative features of the WTDs. However, upon closer inspection it becomes clear that the agreement is not perfect. This is an indication that the renewal assumption is not correct and that subsequent electron waiting times are correlated. With this in mind, we go on to develop a theory of joint WTDs.

\section{Generalized idle time probability}
\label{sec:generalizeditp}
We now generalize the concepts from the last section to an arbitrary number of channels and to an arbitrary number of successive waiting times. To this end, we introduce a generalized ITP from which we obtain the joint WTDs as well as other distributions. In the following section we illustrate our formalism with specific examples.

We consider a scatterer which is connected to $N_i$ incoming and $N_o$ outgoing channels. The generalized ITP is the probability that no particles are detected in any of the outgoing channels during the channel-dependent time intervals $[t^s_{\alpha},t^e_{\alpha}]$. The generalized ITP reads
\begin{equation}
  \Pi(t^s_1, t^e_1; \dots;t^s_{N_o}, t^e_{N_o}) = \left \langle : e^{-\sum_{\alpha=1}^{N_o} \widehat{Q}_\alpha } : \right \rangle,
\label{eq:mcitp}
\end{equation}
having defined projectors for each channel
\begin{equation}
  \widehat{Q}_\alpha = \int_{t^s_\alpha}^{t^e_\alpha} \hat{b}_\alpha^{\dagger}(x) \hat{b}_\alpha(x) dx.
  \label{eq:Qalpha}
\end{equation}
The evaluation of the generalized ITP is discussed in Appendix \ref{sec:computeitp}. The introduction of individual starting and ending times for each channel allows us to compute a variety of WTDs. In each channel, the idle time interval $[t^s_{\alpha},t^e_{\alpha}]$ can be modified and detection events can be inserted by differentiation with respect to the time arguments. The single-channel ITP can always be recovered by letting the length of the intervals in all other channels go to zero
\begin{equation}
\label{eq:singlepi}
\Pi(t_\alpha^s,t_\alpha^e)=\Pi(t_\alpha^s,t_\alpha^e;\{t_{i\neq\alpha}^{s}=t_{i\neq\alpha}^{e}\}),
\end{equation}
where the curly brackets imply $ \forall~i$. {\revision The operators $\widehat{Q}_\alpha$ in
Eq.~\eqref{eq:Qalpha} count the number of particles in the spatial intervals
$[t^s_\alpha,t^e_\alpha]$. For this reason, the generalized ITP is closely related to the joint
particle number statistics in spatial subregions.\cite{shelankov:2008,rammer:2012}}

\subsection{Multiple channels}
\label{sec:multiplechannels}
For the sake of simplicity, we consider only two outgoing channels labeled as $A$ and $B$. The generalization to more channels is straightforward. With two outgoing channels, the ITP has four time arguments $\Pi(t^s_A,t^e_A;t^s_B,t^e_B)$. We consider two different types of WTDs: The two-channel WTD is the distribution of waiting times between detections in either of the two channels. The cross-channel WTD is the distribution of waiting times between a detection in one channel and the next detection in the other channel. These WTDs generally have two time arguments but become one-time quantities for driven systems by averaging over a period following Eq.~\eqref{eq:wtdaveraged}.

We first discuss the two-channel WTD. This is the conditional probability density of detecting an electron at time $t^e$ in either channel, given that the last prior detection happened at the earlier time $t^s$ in any of the two channels. This WTD follows from the generalized ITP by differentiation with respect to the time arguments
\begin{equation}
\label{eq:wtdall}
I(t^s) \mathcal{W}_{AB}(t^s,t^e) = - \partial_{t^s} \partial_{t^e} \Pi(t^s,t^e;t^s,t^e),
\end{equation}
where we have set $t^s=t^s_A=t^s_B$ and $t^e=t^e_A=t^e_B$, since we do not differentiate between the two channels, and $I=I_A+I_B$ is the sum of the particle currents in each channel. If the two channels are uncorrelated, the ITP factorizes as\cite{haack:2014}
\begin{equation}
\label{eq:itpuncorr}
\Pi^{uc}(t^s_A,t^e_A;t^s_B,t^e_B)=\Pi_A(t^s_A,t^e_A)\Pi_B(t^s_B,t^e_B),
\end{equation}
where $\Pi_\alpha$ is the ITP in channel $\alpha$ and the superscript $uc$ stands for uncorrelated. In this case, the two-channel WTD takes on a particularly illuminating form
\begin{equation}
\label{eq:wtduncorr}
\begin{split}
I(t^s) \mathcal{W}^{uc}_{AB}(t^s,t^e) &=I_A(t^s) \mathcal{W}_A(t^s,t^e)\Pi_B(t^s,t^e)\\
&+\mathcal{F}_B(t^s,t^e)\bar{\mathcal{F}}_A(t^s,t^e)+ A\leftrightarrow B.
\end{split}
\end{equation}
The first term represents contributions where both detections happen in channel $A$, while no detections occur in channel $B$. The second term corresponds to contributions where the first detection happens in channel $A$ at time $t^s$ and the next detection occurs in channel $B$ at time $t^e$. Finally, the term $A\leftrightarrow B$ indicates that the roles of the two channels can be interchanged.

In contrast to the single-channel WTD, the Pauli principle does not force the two-channel WTD to vanish at short times (detections in the two channels can occur simultaneously). Evaluating Eq.~\eqref{eq:wtdall} at $t^s=t^e$, we find
\begin{equation}
\label{eq:liftingcorr}
\mathcal{W}_{AB}(t^s,t^s)= 2\frac{\langle  \hat{I}_A(t^s)\hat{I}_B(t^s)  \rangle}{I(t^s)},
\end{equation}
showing that the two-channel WTD at $t^s=t^e$ is  (twice) the coincidence rate in the two channels divided by the total particle current at $t^s$. The factor of two accounts for the fact that either of the two detections can be interpreted as being the first one. For two uncorrelated channels, the expectation value factorizes and we find
\begin{equation}
\label{eq:lifting}
\mathcal{W}^{uc}_{AB}(t^s,t^s)=2\frac{I_A(t^s)I_B(t^s)}{I(t^s)},
\end{equation}
where $I_\alpha$ is the average particle current in channel $\alpha$.

Next we discuss the cross-channel WTD. This is the conditional probability density of waiting until $t^e$ for the first detection in channel $B$ to happen after a
detection has occurred in channel $A$ at the earlier time $t^s$. We note that additional detection may occur in channel
$A$ after the time $t^s$. The cross-channel WTD follows from the ITP as
\begin{equation}
  \label{eq:wait2cc}
  \begin{aligned}
    I_A(t^s)&\mathcal{W}_{A\rightarrow
      B}(t^s,t^e)=-\left.\partial_{t_A^s}\partial_{t^e}\Pi(t_A^s,t^s;t^s,t^e)\right|_{t_A^s=t^s}\\
    &=\left<\hat{b}^\dag_A(t^s)\hat{b}^\dag_B(t^e):e^{-\widehat{Q}_B}:\hat{b}_B(t^e)\hat{b}_A(t^s)\right>.
  \end{aligned}
\end{equation}
We set $t_A^s=t_A^e=t^s$ after taking the derivatives since additional detections in channel $A$ may occur following the first detection. For two uncorrelated channels, we recover the first passage time distribution in channel $B$ since the detection in channel $A$ only defines $t^s$ without influencing channel $B$
\begin{equation}
\label{eq:wait2ccuc}
\mathcal{W}_{A\rightarrow B}^{uc}(t^s,t^e)=\mathcal{F}_B(t^s,t^e).
\end{equation}

For $t^s=t^e$, we again obtain the coincidence rate in the two channels. However, this time without the factor of two since the event in channel $A$ by definition is the first one
\begin{equation}
  \label{eq:liftingcross}
  \mathcal{W}_{A\rightarrow B}(t^s,t^s)=\frac{\langle  \hat{I}_A(t^s)\hat{I}_B(t^s)  \rangle}{I_A(t^s)}.
\end{equation}
For uncorrelated channels this expression reduces to $I_B(t^s)$ in accordance with Eq.~\eqref{eq:wait2ccuc}.

In the following section, we evaluate the two WTDs and show how waiting times may be correlated.

\subsection{Multiple times}
\label{sec:multipletimes}

We are now in position to formulate our theory of joint WTDs. The joint distribution of $n$ waiting times is the conditional probability to find a given sequence of $n$ waiting times. As we will see, the joint WTD can be obtained from the multichannel ITP defined in Eq.~\eqref{eq:mcitp} by introducing auxiliary channels. For the sake of brevity, we consider the case of just two successive waiting times in a single channel. The extension to several channels or waiting times is straightforward.

To find the joint WTD, we consider the two-channel ITP $\Pi(t_\alpha^s,t_\alpha^e;t_\beta^s,t_\beta^e)$, where $\beta$ denotes an auxiliary channel. Eventually, we set $t_\beta^s=t_\beta^e=t^m$ and $\alpha=\beta$ and skip the channel index. Specifically, we express the joint WTD as
\begin{equation}
\begin{aligned}
\label{eq:wait2t}
I(t^s)&\mathcal{W}(t^s,t^m,t^e)=\partial_{t_\alpha^s}\partial_{t_\alpha^e}\partial_{t^e_\beta}\left.\Pi(t_\alpha^s,t_\alpha^e;t_\beta^s,t_\beta^e)\right|_{t_\beta^s=t_\beta^e=t^m}\\
&=\left<\hat{b}^\dag(t^s)\hat{b}^\dag(t^m)\hat{b}^\dag(t^e):e^{-\widehat{Q}}:\hat{b}(t^e)\hat{b}(t^m)\hat{b}(t^s)\right>,
\end{aligned}
\end{equation}
where the operator $\widehat{Q}$ is given in Eq.~\eqref{eq:Qtste}.  Using the auxiliary channel a detection event is inserted at time $t^m$ between the starting $t^s$ and the end time $t^e$ of the interval $[t^s,t^e]$. In a similar way, additional detection events within the interval can be introduced. Equation (\ref{eq:wait2t}) is a central result for the joint distribution of two successive electron waiting times in a single conduction channel. Based on this expression we calculated the joint WTD in Fig.~\ref{fig:wignerdyson} and we will make further use of it in the following section, where we illustrate our method with specific examples.

For a driven conductor, a two-time WTD is obtained by integrating over the period $\mathcal{T}$, cf.~Eq.~\eqref{eq:wtdaveraged},
\begin{equation}
  \label{eq:wtd2taveraged}
  \mathcal{W}(\tau_1, \tau_2) = \frac{\langle \tau \rangle}{\mathcal{T}} \int_0^\mathcal{T} I(t^s)
  \mathcal{W}(t^s,t^s+\tau_1,t^s+\tau_1+\tau_2) dt^s.
  \nonumber
\end{equation}
In addition, we recover the standard WTD by integrating over the time at which the last detection event occurred
\begin{equation}
\label{eq:recwtd}
\mathcal{W}(t^s,t^m)=\int\limits_{t^m}^{\infty}dt^e\mathcal{W}(t^s,t^m,t^e).
\end{equation}

\begin{figure*}
  \centering
  \includegraphics[width=0.9\textwidth]{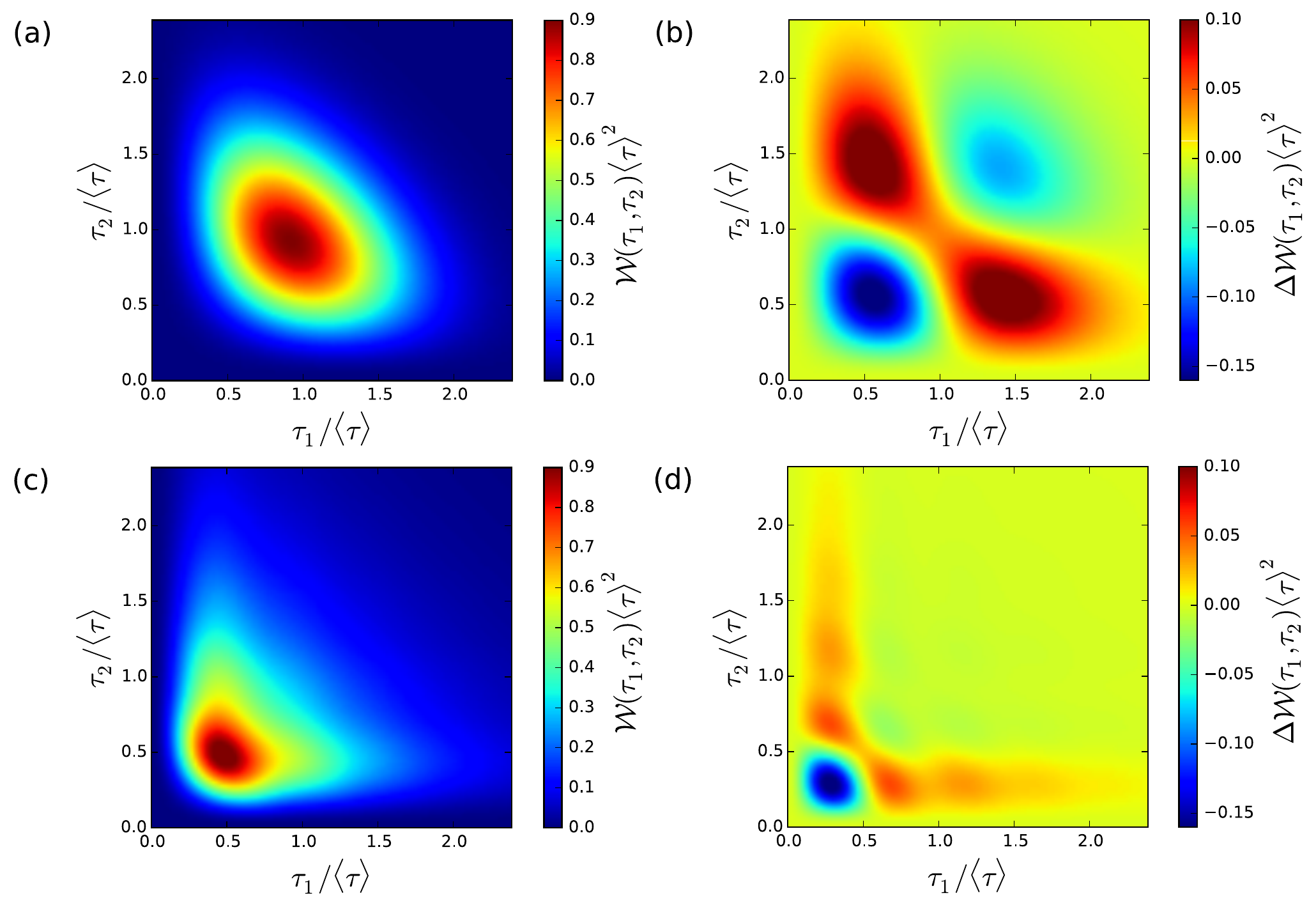}
  \caption{(Color online) Joint waiting time distributions for a static voltage. (a)~Results for the voltage $V_1(t)=V$ applied to contact 1 and no voltage applied to contact 2, $V_2(t)=0$.  The QPC is fully transmitting $T=1$ and we consider the electrons that are transmitted into outgoing channel $B$. The time is given in units of $\langle\tau\rangle=h/(TeV)$. (b) Difference $\Delta \mathcal{W}(\tau_1,\tau_2) = \mathcal{W}(\tau_1,\tau_2) - \mathcal{W}(\tau_1) \mathcal{W}(\tau_2)$ between exact results and results for uncorrelated waiting times. Positive correlations are indicated with red, while areas of negative correlations are blue. (c,d) Similar results for a half-transmitting QPC, $T=1/2$.}
  \label{fig:hbttwotimes}
\end{figure*}

As discussed in Sec.~\ref{sec:renewal}, the joint WTD appears in the expansion of the  WTD for a QPC. Returning to the results in Fig.~\ref{fig:wtdfig1}, the ITP reads
\begin{equation}
\label{eq:itpqpc}
\Pi(t^s,t^e)=\left<:e^{-T\widehat{Q}_1}:\right>=\left<:e^{(R-1)\widehat{Q}_1}:\right>,
\end{equation}
where $\widehat{Q}_1$ in Eq.~\eqref{eq:qops} acts on the incoming channel $1$. We can formally expand this expression in either $T$ or $R$.\cite{albert:2012,dasenbrook:2014} To zeroth order in $T$, the ITP is unity as no electrons are transmitted through the QPC. The $n$'th order term in the expansion yields the reduction of the ITP due to the probability that $n$ particles were transmitted through the QPC. To zeroth order in $R$, the ITP equals the ITP for a fully transmitting QPC. The $n$'th order term in the expansion in $R$ equals the increase in the ITP due to the probability that $n$ particles were reflected.

Here we perform an expansion in $R$, and by differentiating the ITP with respect to $t^s$ and $t^e$ we obtain
\begin{equation}
\mathcal{W}(t^s,t^e)=T\sum\limits_{n=0}^{\infty}R^n\mathcal{W}^{(n)}(t^s,t^e)
\end{equation}
where
\begin{equation}
\mathcal{W}^{(n)}(t^s,t^e)=\frac{1}{n!}\int\limits_{t^s}^{t^e}dt^1\cdots dt^n\mathcal{W}_{\mathrm{in}}(t^s,t^1,\cdots,t^n,t^e)
\end{equation}
is the WTD given that $n$ reflections on the QPC have occurred and $\mathcal{W}_{\mathrm{in}}(t^s,t^1,\cdots,t^n,t^e)$ is the joint WTD for $n+2$ detection events in the incoming channel $1$. Each term in the expansion is the probability density for $n$ particles to be reflected (corresponding to the prefactor $R^n$) followed by a transmission (corresponding to the prefactor $T$). We integrate over all possible times that the reflections can occur and the factor $1/n!$ corrects for multiple counting of reflections. We see that joint WTDs occur already in the expansion of the WTD for a QPC. By averaging over $t^s$ and making a renewal assumption, we can recover the results from Sec.~\ref{sec:renewal}.

\section{Correlated waiting times}
\label{sec:correlationsinwaitingtimes}

We now illustrate our formalism for joint WTDs using the setup in Fig.~\ref{fig:setup}. First we consider the partitioning of electrons emitted from one source. In the second example we discuss an electronic analog of the Hong-Ou-Mandel interferometer, where electrons from different input channels interfere on the QPC. \cite{liu:1998,burkard:2000,olkhovskaya:2008,giovannetti:2006,jonckheere:2012,bocquillon:2013}

\subsection{Single-source partitioning}
\label{sec:singlesourcepartitioning}

\begin{figure*}
  \centering
  \includegraphics[width=0.95\textwidth]{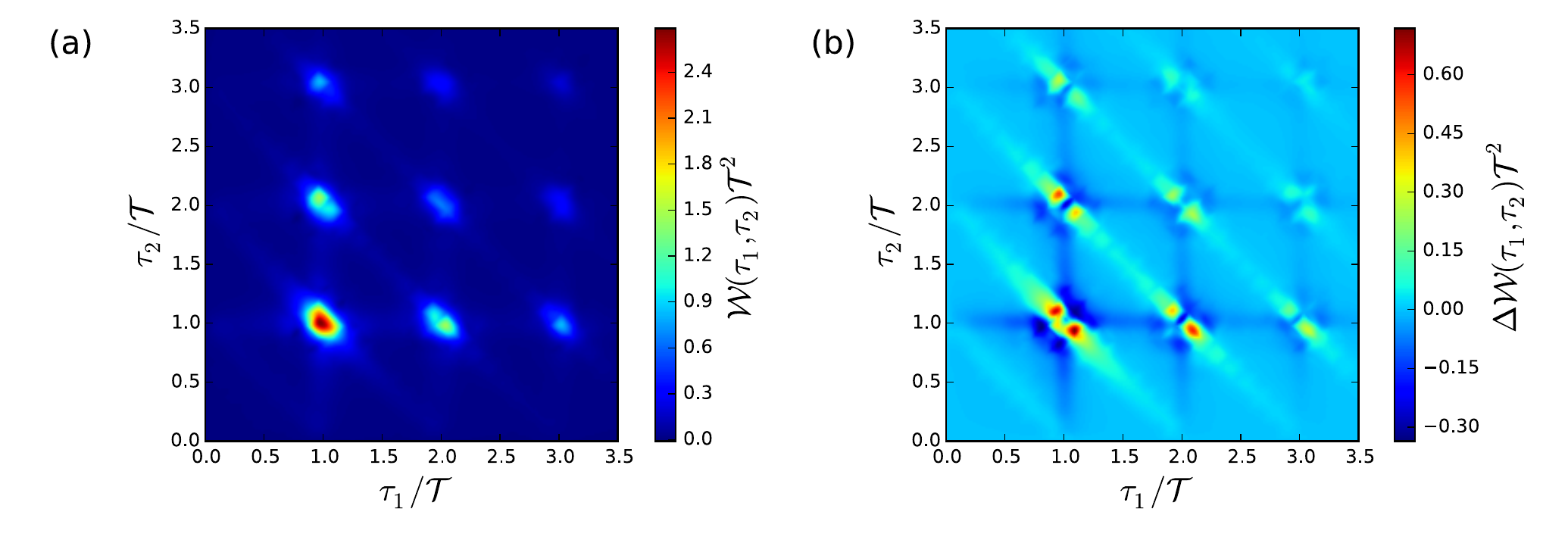}
  \caption{(Color online) Joint waiting time distribution for levitons. (a)~Results for a series of Lorentzian-shaped voltage pulses of unit charge applied to contact 1. The width of the pulses is $\Gamma=0.05\mathcal{T}$ and the QPC is tuned to $T=1/2$. The time is given in units of the period of the pulses~$\mathcal{T}$. (b) Difference $\Delta \mathcal{W}(\tau_1,\tau_2) = \mathcal{W}(\tau_1,\tau_2) - \mathcal{W}(\tau_1) \mathcal{W}(\tau_2)$ between exact results and results for uncorrelated waiting times. Positive correlations are indicated with red, while areas of negative correlations are blue.}
  \label{fig:hbttwotimeslevitons}
\end{figure*}

We consider the setup in Fig.~\ref{fig:setup} with contact $2$ grounded and contact $1$ biased with a constant voltage or a train of Lorentzian-shaped voltage pulses. Figures~\ref{fig:itpfptdwtd} and \ref{fig:wtdfig1} show single-channel WTDs for this setup. We now go on to calculate two-channel and joint WTDs.

The two-channel WTD is the distribution of waiting times between successive electrons irrespective of the channel in which they are detected. As the QPC merely distributes the incoming electrons into the two outgoing channels, we expect simply to recover the WTD of the incoming electrons from channel 1. Indeed, using our formalism we find from Eq.~\eqref{eq:wtdall} that
\begin{equation}
\begin{split}
  I(t^s) \mathcal{W}_{AB}(t^s,t^e)& = - \partial_{t^s} \partial_{t^e}\left \langle : e^{
      -\widehat{Q}_A - \widehat{Q}_B} : \right \rangle\\
  &=  - \partial_{t^s} \partial_{t^e}\left \langle : e^{-\widehat{Q}_1}: \right \rangle,
\end{split}
\label{eq:hbt2c}
\end{equation}
where the integrations in the $\widehat{Q}$ operators run from $t^s$ to $t^e$ and $I(t^s)$ is the average current in channel $1$. In addition, we have used Eqs.~\eqref{eq:qops} and \eqref{eq:qops2} to relate $\widehat{Q}_\alpha$ to $\widehat{Q}_1$ combined with Eq.~\eqref{eq:propcons}. For a constant voltage, the WTD was evaluated in Ref.~\onlinecite{albert:2012} and found to be well-approximated by a Wigner-Dyson distribution reflecting Fermi correlations between the incoming electrons. For a train of Lorentzian-shaped voltage pulses, the WTD was evaluated in Ref.~\onlinecite{dasenbrook:2014} and found to be peaked around the period of the driving with small satellite peaks due to the finite overlap of the voltage pulses. As we see here, the QPC  has no effect on the WTD if we consider detections irrespective of the channel where they happen. This argument only applies if the QPC partitions electrons emitted by a single source. In that case, detections in the outgoing channels are not statistically independent, and the suppression of the WTD at $\tau=0$ remains.

Next, we consider the cross-channel WTD. This is the distribution of waiting times between a detection in one outgoing channel and the next detection in the other channel. Since the  QPC just randomly partitions the incoming electrons into the outgoing channels, we expect the cross-channel WTD to equal the single-channel WTD. The cross-channel WTD is conditioned on the first detection happening in channel $A$, whereas the single-channel WTD is conditioned on the first detection happening in channel $B$. In either case,
the reflection or transmission of the first particle does not influence the particles that traverse the QPC at a later time.  Indeed, within our formalism we readily find, cf.~Eq.~\eqref{eq:wait2cc},
\begin{align}
  \label{eq:crosswtdhbt}
  &\mathcal{W}_{A\rightarrow B}(t^s,t^e)  = \mathcal{W}_B(t^s,t^e) \nonumber\\
  &= \frac{RT}{I_A(t^s)} \left \langle \hat{a}_1^\dagger(t^s) \hat{a}_1^\dagger(t^e) : e^{- T\widehat{Q}_1} : \hat{a}_1(t^e) \hat{a}_1(t^s) \right \rangle,
\end{align}
having used $I_A(t^s)/R =I_B(t^s)/T= I(t^s)$.

For the partitioning of electrons emitted from a single source, we see that the two-channel WTD and the cross-channel WTD do not provide additional information compared with the WTD itself. This is a direct consequence of the operator proportionality $\widehat{Q}_A\propto \widehat{Q}_B$ and it does not hold when both contacts emit electrons above the Fermi sea as we will see in the following example.

We now consider just one of the outgoing channels and calculate the joint probability distribution of finding two successive waiting times $\tau_1$ and $\tau_2$ using Eq.~\eqref{eq:wait2t}. For a constant voltage $V$, the joint WTD at full transmission is shown in Fig.~\ref{fig:wignerdyson}. As mentioned in the introduction, the joint WTD is well-approximated by the generalized Wigner-Dyson distribution in Eq.~\eqref{eq:2WD}. We find that the joint WTD is symmetric with
respect to an exchange of the waiting times, $\tau_1 \leftrightarrow \tau_2$. This symmetry implies that the WTD does not
change if we invert the spatial arguments of all the $\hat{b}$ operators in Eq.~\eqref{eq:wait2t}. The symmetry in the WTD is thus a consequence of the spatial inversion symmetry of the many-body wavefunction.

In Fig.~\ref{fig:hbttwotimes}, we show joint WTDs both for full transmission and at half transmission. To highlight possible correlations, we also show the difference
\begin{equation}
\Delta \mathcal{W}(\tau_1,\tau_2) = \mathcal{W}(\tau_1,\tau_2) - \mathcal{W}(\tau_1) \mathcal{W}(\tau_2),
\end{equation}
between the joint WTD and a factorized WTD corresponding to uncorrelated waiting times.\cite{cox:1962} The figure clearly demonstrates that electron waiting times are correlated.
The probability of observing a waiting time which is shorter (longer) than the mean waiting time $\langle \tau \rangle$ is reduced if the previous waiting time was already shorter (longer) than the mean waiting time. On the other hand,  a short waiting time will likely be followed by a long waiting time and vice versa. This conclusion holds for both values of the transmission.

A similar analysis can be carried out for levitons emitted by a train of Lorentzian-shaped voltage pulses. In Fig.~\ref{fig:hbttwotimeslevitons} we show $\mathcal{W}(\tau_1,\tau_2)$ and $\Delta \mathcal{W}(\tau_1,\tau_2)$ for levitons incident on a QPC tuned to half transmission. Again, we observe a symmetry under the interchange of waiting times due to the spatial inversion symmetry of the fermionic many-body state. The joint WTD displays a lattice-like structure in contrast to the results for the dc-biased source in Fig.~\ref{fig:hbttwotimes}. This reflects the regular on-demand nature of the leviton source. Also here, the probability to observe two short or two long waiting times after each other is suppressed, while a short (long) waiting time is likely followed by a long (short) waiting time.

{\revision Due to the external driving, the sum of two subsequent waiting times is likely to equal a
multiple of the period. If a particular waiting time is shorter than the average waiting time, this
will not affect the absolute emission time of the next electron. Consequently, the next waiting time
will most likely be longer. Thus, a strong regularity in the absolute electron emission times leads
to strong dependences of the waiting times on each other. For the case that the voltage pulses
overlap more and more until the limit of a constant voltage is reached, the regularity in the
emission times is provided by the Pauli principle. Moreover, this effect is independent of the QPC
transmission. In the low transmission regime (not shown), however, electron emission becomes
increasingly rare and transport resembles a Poisson process.\cite{albert:2012} Correlations between
waiting times will then eventually be negligible.}

\subsection{Hong-Ou-Mandel interferometry}
\label{sec:hom}

As a second application, we consider the electronic analog of the Hong-Ou-Mandel experiment developed in quantum optics.\cite{liu:1998,burkard:2000,olkhovskaya:2008,giovannetti:2006,jonckheere:2012,bocquillon:2013} In this case, a driven on-demand source in each incoming channel emits single electrons onto the QPC in Fig.~\ref{fig:setup} with a possible time delay $\Delta \tau$ between the emissions from the two sources. Such experiments have recently been realized with two mesoscopic capacitors\cite{bocquillon:2013} and two leviton sources.\cite{dubois:2013} Here we treat the periodic emission of levitons onto the QPC.

\begin{figure*}
  \centering
  \includegraphics[width=0.9\textwidth]{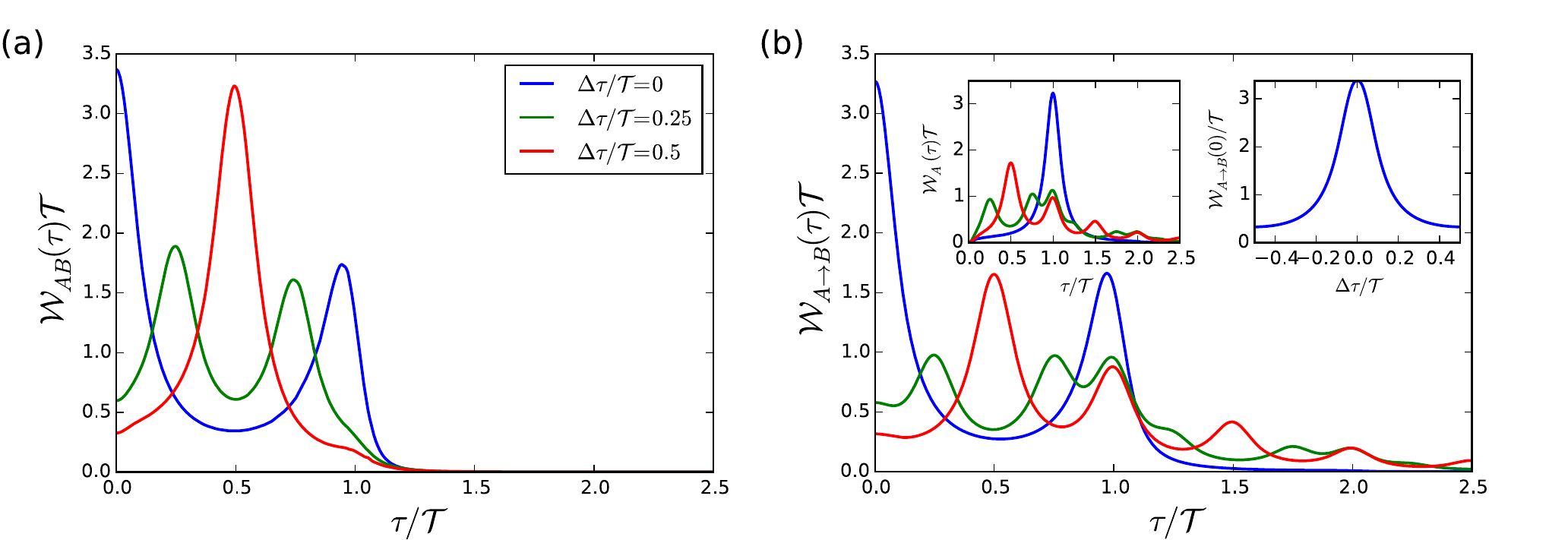}
  \caption{(Color online) Distributions of waiting times for a Hong-Ou-Mandel experiment with levitons. Two sources emit levitons of pulse width $\Gamma=0.05\mathcal{T}$ toward a QPC with transmission $T=1/2$. The tunable time delay between the sources is denoted as $\Delta \tau$. (a) WTD for waiting times between events occurring in either of the two outgoing channels for different values of the time delay. (b) WTD for waiting times between detection events in different channels. The left inset shows the WTD for just a single channel. The right inset shows the WTD at $\tau=0$ as a function of the time delay $\Delta \tau$ between the sources.}
  \label{fig:hom}
\end{figure*}

Figure~\ref{fig:hom}a shows the distribution of waiting times between detection events in either of the two outgoing channels. With zero time delay, incoming levitons antibunch on the QPC and there is a large probability of detecting the two outgoing levitons nearly simultaneously. The peak in the WTD around the period $\mathcal{T}$ corresponds to the waiting time between the last leviton in one cycle and the first one in the next cycle. Since the probability of measuring the waiting time between two levitons within one cycle is equal to the probability of measuring the waiting time between levitons in different cycles, the areas underneath the two peaks are the same.

In general, for a given time delay $\Delta \tau$, we find peaks in the WTD at $\Delta \tau$ and at $\mathcal{T} - \Delta \tau$, corresponding to the waiting times between levitons within the same cycle and the waiting between levitons in different cycles. In the special case $\Delta \tau = \mathcal{T} /2$, the two peaks merge into one, except for a small remaining feature at $\tau = \mathcal{T}$. This feature is related to the satellite peak seen in the WTD for levitons in a single channel with perfect transmission.\cite{dasenbrook:2014}  The WTD decays strongly for waiting times beyond the period, since all levitons will be detected independently of the QPC. This reflects the high reliability of the leviton sources which emit exactly one electron per cycle. This example demonstrates how the two-channel WTD provides information about the synchronization of the two sources. Importantly, this detailed characterization does not depend on the transmission of the QPC.

In Fig.~\ref{fig:hom}b we turn to the distribution of waiting times between detections in different channels. This distribution is rather similar to the WTD for a single channel, shown in the left inset for comparison. Unlike the single-channel WTD, the cross-channel WTD is not suppressed to zero at $\tau=0$, since detections can occur simultaneously in the two channels. For finite delay times between the two sources, the particles can go into the same outgoing channel, and both WTDs show a peaked structure even for waiting times larger than the period.

Due to the Pauli principle, two levitons arriving simultaneously at the QPC will go into different outgoing channels with unit probability, independently of the transmission. Thus, for zero time delay all results are independent of the QPC transmission $T$ and reduce to the results for full transmission.  As the time delay between the two sources is increased, the overlap between levitons arriving at the QPC decreases, and simultaneous detections in the outgoing channels become increasingly rare. For $\Delta \tau = \mathcal{T}/4$, the cross-channel WTD essentially shows four peaks within the first period. The peak at $\tau = 0$ is a relict of the fermionic antibunching due to the
finite overlap of the pulses. The peaks at $\Delta \tau$ and at $\mathcal{T} - \Delta \tau$ are caused by two successive particles entering opposite arms of the interferometer, while the peak at $\mathcal{T}$ corresponds to two successive particles entering the same arm and the next particle entering the opposite arm. For the maximal detuning $\Delta \tau  =  \mathcal{T}/2$, the WTD resembles
that of just a single channel.

Due to the Pauli principle the cross-channel WTD of a single source is suppressed for zero waiting times. By constrast, with two sources we observe a local maximum due to the antibunching of electrons that arrive simultaneously at the QPC. The right inset of Fig.~\ref{fig:hom}b shows the cross-channel WTD at $\tau=0$ as a function of the time delay, with a maximum for zero time delay in analogy with the Hong-Ou-Mandel {\revision peak} found in the zero-frequency current cross-correlations.\cite{jonckheere:2012,bocquillon:2013}

We see that the two-channel and cross-channel WTDs contain different information about the emission processes.  The two-channel WTD mostly contains information about the sources alone, in particular their synchronization. The cross-channel WTD contains additional information about the QPC and shows more prominent signatures of fermionic antibunching.

\section{Conclusions}
\label{sec:conclusions}
We have developed a general framework for calculating joint WTDs in electronic multi-channel conductors. The central building block of our formalism is the generalized idle time probability, i.~e.~the joint probability for no detections to occur in the outgoing channels. By calculating the joint WTD for a single conduction channel, we have explicitly demonstrated that the electron waiting times in coherent conductors are correlated due to the fermionic statistics encoded in the many-body state.

Drawing on the analogy between random matrices and free fermions, we have shown that the joint WTD for a fully transmitting conduction channel is well-approximated by a generalized Wigner-Dyson distribution.  In contrast to a renewal process with uncorrelated waiting times, we find that the probability of observing a long (short) waiting time following a short (long) waiting time is increased, while finding two long or two short waiting times in succession is less likely. This holds both for electrons coming from a dc-biased contact and for levitons emitted on top of the Fermi sea by applying Lorentzian-shaped voltage pulses to the contact.

Correlations between electrons in different outgoing channels also show up in the distributions of waiting times between detections in different channels. We have defined multi-channel and cross-channel WTDs and illustrated these concepts for a QPC in a chiral setup where electrons are injected in either one or both incoming channels. In a fermionic Hong-Ou-Mandel experiment, where  electrons interfere on the QPC, the two-channel WTD provides information about the synchronization of the  sources, while the cross-channel WTD shows signatures of the scatterer and the fermionic antibunching.

Our formalism relies on the ability to detect single electrons above the Fermi level. A quantum theory of such a detector is currently being developed. It would also be interesting to make use of WTDs to detect entanglement in electronic conductors, for instance by formulating a set of Leggett-Garg inequalities for the joint WTDs.

\begin{acknowledgments}
We thank M.~Albert and G.~Haack for many useful discussions. CF is affiliated with Centre for Quantum Engineering at Aalto University. PH gratefully acknowledges the hospitality of McGill University where part of the work was done. The work was supported by the Swiss NSF.
\end{acknowledgments}

\appendix

\section{Computation of the generalized ITP}
\label{sec:computeitp}
To compute the generalized ITP in Eq.~\eqref{eq:mcitp}, we map the outgoing operators onto the incoming ones using the Floquet scattering matrix\cite{moskalets:2002,moskalets:book}
\begin{equation}
  \label{eq:Smatrix}
  \hat{b}_\alpha(E) = \sum_{n=-\infty}^{\infty} \sum_{\beta=1}^{N_i}
  S_{F,\alpha\beta}(E,E_n)\hat{a}_\beta(E_n),
\end{equation}
where $\Omega$ is the frequency of the external driving and we have defined the energies
\begin{equation}
E_n = E + n\hbar \Omega,
\end{equation}
for integer $n$. Since the incoming equilibrium particles are described by a Slater determinant, we may evaluate the quantum statistical average in Eq.~\eqref{eq:mcitp} and we thereby obtain a determinant formula\cite{albert:2012,dasenbrook:2014,haack:2014} of the form
\begin{equation}
  \Pi(t^s_1, t^e_1; \dots; t^s_{N_o}, t^e_{N_o}) = \det(\mathbb{I} - \mathcal{Q}(t^s_1, t^e_1; \dots; t^s_{N_o}, t^e_{N_o})). \nonumber
\end{equation}
Here $\mathbb{I}$ is the identity matrix and $\mathcal{Q}$ contains the single-particle matrix elements of the operator $\sum_\alpha \widehat{Q}_\alpha$.

The matrix $\mathcal{Q}$ has a block form with elements in the (incoming) channel space
\begin{equation}
  \label{eq:Qmatrixblockform}
  [\mathcal{Q}(t^s_1, t^e_1, \dots, t^s_{N_o}, t^e_{N_o})]_{\alpha \beta} = \mathbf{K}_{\alpha \beta}
\end{equation}
with the $\mathbf{K}_{\alpha \beta}$ being matrices in energy space. In the general case of a time-dependent scatterer with many channels described by the Floquet scattering matrix in Eq.~\eqref{eq:Smatrix}, these matrices have the elements
\begin{equation}
\begin{split}
  \label{eq:Kelements}
  [\mathbf{K}_{\alpha \beta}]_{E,E^{'}} &= \sum_{\gamma=1}^{N_o} \mathop{\mathop{\sum^\infty
      \sum^\infty}_{m=-\lfloor E/\hbar \Omega \rfloor}}_{n=-\lfloor E^{'}/\hbar \Omega
    \rfloor} S^\dagger_{F,\gamma \alpha}(E_m,E) S_{F,\gamma \beta}(E^{'}_n,E^{'}) \\
  &\quad \times \Theta(- E) \Theta(- E^{'})
  [K(t^s_\gamma,t^e_\gamma)]_{E_m-E^{'}_n},
\end{split}
\end{equation}
where we have defined
\begin{equation}
[K(t^s_\gamma,t^e_\gamma)]_E = \frac{\kappa}{\pi} e^{-i E (t^s_\gamma+t^e_\gamma)/2}\frac{\sin(E(t^e_\gamma-t^s_\gamma)/2)}{E}.
\end{equation}
In addition, the floor function is denoted as $\lfloor \cdot \rfloor$ and we have discretized\cite{hassler:2008,albert:2012,haack:2014} the transport window $[E_F,E_F+eV]$ into compartments of width $\kappa=eV/\mathcal{N}$, where $\mathcal{N}$ is the number of particles. We always consider the limit $\mathcal{N} \to \infty$. The matrix has a block form with respect to the incoming channels, while we sum over the indices of the outgoing channels. The result in Eq.~\eqref{eq:Kelements} generalizes the earlier expression for a single channel\cite{dasenbrook:2014} to many  channels.

\section{Levitons}
\label{app:levitons}
Here we briefly discuss the Floquet scattering theory for the periodic creation of levitons.\cite{ivanov:1997,levitov:1996,keeling:2006,dubois:2013prb} Applying a series of periodic Lorentzian voltage pulses of unit
charge,
\begin{equation}
  \label{eq:Vtlorentzians}
  eV(t)=\sum\limits_{j=-\infty}^{\infty}\frac{2\hbar\Gamma}{\left(t-t_0-j\mathcal{T}\right)^2+\Gamma^2},
\end{equation}
to one of the reservoirs leads to the creation of one clean electron excitation per period --- a leviton --- without accompanying electron-hole pairs. Here, the period of the pulses is denoted as $\mathcal{T}=2\pi/\Omega$ and the parameter $t_0$ can be adjusted to shift the pulses in time. In the adiabatic limit, where the variation of the voltage is slow compared to the scattering time, the elements of the Floquet scattering matrix can be obtained by Fourier transforming the time dependent scattering phase in Eq.~\eqref{eq:smatvolt} with
$V(t)$ defined in Eq.~\eqref{eq:Vtlorentzians}. We then obtain\cite{dubois:2013prb}
\begin{equation}
  \label{eq:levitonfloquetsmatrix}
  S_n = e^{in\Omega t_0} \begin{cases}
    -2 e^{-n\Omega\Gamma}\sinh(\Omega\Gamma)  & \text{if } n > 0 \\
    e^{-\Omega\Gamma} & \text{if } n = 0 \\
    0 & \text{otherwise}
  \end{cases}.
\end{equation}
This Floquet scattering matrix relates particles incident on the scatterer to equilibrium particles through Eq.~\eqref{eq:floquetSmatrix}. Together with the scattering matrix of the scatterer itself, it yields the Floquet scattering matrix used for the calculation of the generalized idle time probability in Eq.~\eqref{eq:Kelements}. For the Hong-Ou-Mandel calculations, the
time delay between the sources can be tuned by changing the parameter $t_0$ in one of the matrix elements in the channel space.


\begin{thebibliography}{61}%
\makeatletter
\providecommand \@ifxundefined [1]{%
 \@ifx{#1\undefined}
}%
\providecommand \@ifnum [1]{%
 \ifnum #1\expandafter \@firstoftwo
 \else \expandafter \@secondoftwo
 \fi
}%
\providecommand \@ifx [1]{%
 \ifx #1\expandafter \@firstoftwo
 \else \expandafter \@secondoftwo
 \fi
}%
\providecommand \natexlab [1]{#1}%
\providecommand \enquote  [1]{``#1''}%
\providecommand \bibnamefont  [1]{#1}%
\providecommand \bibfnamefont [1]{#1}%
\providecommand \citenamefont [1]{#1}%
\providecommand \href@noop [0]{\@secondoftwo}%
\providecommand \href [0]{\begingroup \@sanitize@url \@href}%
\providecommand \@href[1]{\@@startlink{#1}\@@href}%
\providecommand \@@href[1]{\endgroup#1\@@endlink}%
\providecommand \@sanitize@url [0]{\catcode `\\12\catcode `\$12\catcode
  `\&12\catcode `\#12\catcode `\^12\catcode `\_12\catcode `\%12\relax}%
\providecommand \@@startlink[1]{}%
\providecommand \@@endlink[0]{}%
\providecommand \url  [0]{\begingroup\@sanitize@url \@url }%
\providecommand \@url [1]{\endgroup\@href {#1}{\urlprefix }}%
\providecommand \urlprefix  [0]{URL }%
\providecommand \Eprint [0]{\href }%
\providecommand \doibase [0]{http://dx.doi.org/}%
\providecommand \selectlanguage [0]{\@gobble}%
\providecommand \bibinfo  [0]{\@secondoftwo}%
\providecommand \bibfield  [0]{\@secondoftwo}%
\providecommand \translation [1]{[#1]}%
\providecommand \BibitemOpen [0]{}%
\providecommand \bibitemStop [0]{}%
\providecommand \bibitemNoStop [0]{.\EOS\space}%
\providecommand \EOS [0]{\spacefactor3000\relax}%
\providecommand \BibitemShut  [1]{\csname bibitem#1\endcsname}%
\let\auto@bib@innerbib\@empty
%</preamble>
\bibitem [{\citenamefont {Blanter}\ and\ \citenamefont
  {B\"uttiker}(2000)}]{blanter:2000}%
  \BibitemOpen
  \bibfield  {author} {\bibinfo {author} {\bibfnamefont {Ya.~M.}\ \bibnamefont
  {Blanter}}\ and\ \bibinfo {author} {\bibfnamefont {M.}~\bibnamefont
  {B\"uttiker}},\ }\bibfield  {title} {\enquote {\bibinfo {title} {Shot noise
  in mesoscopic conductors},}\ }\href
  {http://www.sciencedirect.com/science/article/pii/S0370157399001234}
  {\bibfield  {journal} {\bibinfo  {journal} {Phys. Rep.}\ }\textbf {\bibinfo
  {volume} {336}},\ \bibinfo {pages} {1} (\bibinfo {year} {2000})}\BibitemShut
  {NoStop}%
\bibitem [{\citenamefont {Lesovik}(1989)}]{lesovik:1989}%
  \BibitemOpen
  \bibfield  {author} {\bibinfo {author} {\bibfnamefont {G.~B.}\ \bibnamefont
  {Lesovik}},\ }\bibfield  {title} {\enquote {\bibinfo {title} {Excess quantum
  noise in 2D ballistic point contacts},}\ }\href@noop {}
  {\bibfield  {journal} {\bibinfo  {journal} {JETP Lett.}\ }\textbf {\bibinfo
  {volume} {49}},\ \bibinfo {pages} {594} (\bibinfo {year} {1989})}\BibitemShut
  {NoStop}%
\bibitem [{\citenamefont {B\"uttiker}(1990)}]{buttiker:1990}%
  \BibitemOpen
  \bibfield  {author} {\bibinfo {author} {\bibfnamefont {M.}~\bibnamefont
  {B\"uttiker}},\ }\bibfield  {title} {\enquote {\bibinfo {title} {Scattering
  theory of thermal and excess noise in open conductors},}\ }\href {\doibase
  10.1103/PhysRevLett.65.2901} {\bibfield  {journal} {\bibinfo  {journal}
  {Phys. Rev. Lett.}\ }\textbf {\bibinfo {volume} {65}},\ \bibinfo {pages}
  {2901} (\bibinfo {year} {1990})}\BibitemShut {NoStop}%
\bibitem [{\citenamefont {Martin}\ and\ \citenamefont
  {Landauer}(1992)}]{martin:1992}%
  \BibitemOpen
  \bibfield  {author} {\bibinfo {author} {\bibfnamefont {T.}~\bibnamefont
  {Martin}}\ and\ \bibinfo {author} {\bibfnamefont {R.}~\bibnamefont
  {Landauer}},\ }\bibfield  {title} {\enquote {\bibinfo {title} {Wave-packet
  approach to noise in multichannel mesoscopic systems},}\ }\href {\doibase
  10.1103/PhysRevB.45.1742} {\bibfield  {journal} {\bibinfo  {journal} {Phys.
  Rev. B}\ }\textbf {\bibinfo {volume} {45}},\ \bibinfo {pages} {1742}
  (\bibinfo {year} {1992})}\BibitemShut {NoStop}%
\bibitem [{\citenamefont {Kumar}\ \emph {et~al.}(1996)\citenamefont {Kumar},
  \citenamefont {Saminadayar}, \citenamefont {Glattli}, \citenamefont {Jin},\
  and\ \citenamefont {Etienne}}]{kumar:1996}%
  \BibitemOpen
  \bibfield  {author} {\bibinfo {author} {\bibfnamefont {A.}~\bibnamefont
  {Kumar}}, \bibinfo {author} {\bibfnamefont {L.}~\bibnamefont {Saminadayar}},
  \bibinfo {author} {\bibfnamefont {D.~C.}\ \bibnamefont {Glattli}}, \bibinfo
  {author} {\bibfnamefont {Y.}~\bibnamefont {Jin}}, \ and\ \bibinfo {author}
  {\bibfnamefont {B.}~\bibnamefont {Etienne}},\ }\bibfield  {title} {\enquote
  {\bibinfo {title} {Experimental test of the quantum shot noise reduction
  theory},}\ }\href {\doibase 10.1103/PhysRevLett.76.2778} {\bibfield
  {journal} {\bibinfo  {journal} {Phys. Rev. Lett.}\ }\textbf {\bibinfo
  {volume} {76}},\ \bibinfo {pages} {2778} (\bibinfo {year}
  {1996})}\BibitemShut {NoStop}%
\bibitem [{\citenamefont {Levitov}\ and\ \citenamefont
  {Lesovik}(1993)}]{levitov:1993}%
  \BibitemOpen
  \bibfield  {author} {\bibinfo {author} {\bibfnamefont {L.~S.}\ \bibnamefont
  {Levitov}}\ and\ \bibinfo {author} {\bibfnamefont {G.~B.}\ \bibnamefont
  {Lesovik}},\ }\bibfield  {title} {\enquote {\bibinfo {title} {Charge
  distribution in quantum shot noise},}\ }\href@noop {} {\bibfield  {journal}
  {\bibinfo  {journal} {JETP Lett.}\ }\textbf {\bibinfo {volume} {58}},\
  \bibinfo {pages} {230} (\bibinfo {year} {1993})}\BibitemShut {NoStop}%
\bibitem [{\citenamefont {Levitov}\ \emph {et~al.}(1996)\citenamefont
  {Levitov}, \citenamefont {Lee},\ and\ \citenamefont
  {Lesovik}}]{levitov:1996}%
  \BibitemOpen
  \bibfield  {author} {\bibinfo {author} {\bibfnamefont {L.~S.}\ \bibnamefont
  {Levitov}}, \bibinfo {author} {\bibfnamefont {H.}~\bibnamefont {Lee}}, \ and\
  \bibinfo {author} {\bibfnamefont {G.~B.}\ \bibnamefont {Lesovik}},\
  }\bibfield  {title} {\enquote {\bibinfo {title} {Electron counting statistics
  and coherent states of electric current},}\ }\href {\doibase
  10.1063/1.531672} {\bibfield  {journal} {\bibinfo  {journal} {J. Math.
  Phys.}\ }\textbf {\bibinfo {volume} {37}},\ \bibinfo {pages} {4845} (\bibinfo
  {year} {1996})}\BibitemShut {NoStop}%
\bibitem [{\citenamefont {Nazarov}(2003)}]{nazarov:2003}%
  \BibitemOpen
  \bibinfo {editor} {\bibfnamefont {Yu.~V.}\ \bibnamefont {Nazarov}},\ ed.,\
  \href@noop {} {\emph {\bibinfo {title} {Quantum Noise in Mesoscopic
  Physics}}}\ (\bibinfo  {publisher} {Kluwer, Dordrecht},\ \bibinfo {year}
  {2003})\BibitemShut {NoStop}%
\bibitem [{\citenamefont {Shelankov}\ and\ \citenamefont
  {Rammer}(2003)}]{shelankov:2003}%
  \BibitemOpen
  \bibfield  {author} {\bibinfo {author} {\bibfnamefont {A.}~\bibnamefont
  {Shelankov}}\ and\ \bibinfo {author} {\bibfnamefont {J.}~\bibnamefont
  {Rammer}},\ }\bibfield  {title} {\enquote {\bibinfo {title} {Charge transfer
  counting statistics revisited},}\ }\href@noop {} {\bibfield  {journal}
  {\bibinfo  {journal} {Europhys. Lett.}\ }\textbf {\bibinfo {volume} {63}},\
  \bibinfo {pages} {485} (\bibinfo {year} {2003})}\BibitemShut {NoStop}%
\bibitem [{\citenamefont {Vanevi\ifmmode~\acute{c}\else \'{c}\fi{}}\ \emph
  {et~al.}(2007)\citenamefont {Vanevi\ifmmode~\acute{c}\else \'{c}\fi{}},
  \citenamefont {Nazarov},\ and\ \citenamefont {Belzig}}]{vanevic:2007}%
  \BibitemOpen
  \bibfield  {author} {\bibinfo {author} {\bibfnamefont {M.}~\bibnamefont
  {Vanevi\ifmmode~\acute{c}\else \'{c}\fi{}}}, \bibinfo {author} {\bibfnamefont
  {Yu.~V.}\ \bibnamefont {Nazarov}}, \ and\ \bibinfo {author} {\bibfnamefont
  {W.}~\bibnamefont {Belzig}},\ }\bibfield  {title} {\enquote {\bibinfo {title}
  {Elementary events of electron transfer in a voltage-driven quantum point
  contact},}\ }\href {\doibase 10.1103/PhysRevLett.99.076601} {\bibfield
  {journal} {\bibinfo  {journal} {Phys. Rev. Lett.}\ }\textbf {\bibinfo
  {volume} {99}},\ \bibinfo {pages} {076601} (\bibinfo {year}
  {2007})}\BibitemShut {NoStop}%
\bibitem [{\citenamefont {Vanevi\ifmmode~\acute{c}\else \'{c}\fi{}}\ \emph
  {et~al.}(2008)\citenamefont {Vanevi\ifmmode~\acute{c}\else \'{c}\fi{}},
  \citenamefont {Nazarov},\ and\ \citenamefont {Belzig}}]{vanevic:2008}%
  \BibitemOpen
  \bibfield  {author} {\bibinfo {author} {\bibfnamefont {M.}~\bibnamefont
  {Vanevi\ifmmode~\acute{c}\else \'{c}\fi{}}}, \bibinfo {author} {\bibfnamefont
  {Yu.~V.}\ \bibnamefont {Nazarov}}, \ and\ \bibinfo {author} {\bibfnamefont
  {W.}~\bibnamefont {Belzig}},\ }\bibfield  {title} {\enquote {\bibinfo {title}
  {Elementary charge-transfer processes in mesoscopic conductors},}\ }\href
  {\doibase 10.1103/PhysRevB.78.245308} {\bibfield  {journal} {\bibinfo
  {journal} {Phys. Rev. B}\ }\textbf {\bibinfo {volume} {78}},\ \bibinfo
  {pages} {245308} (\bibinfo {year} {2008})}\BibitemShut {NoStop}%
\bibitem [{\citenamefont {Abanov}\ and\ \citenamefont
  {Ivanov}(2008)}]{abanov:2008}%
  \BibitemOpen
  \bibfield  {author} {\bibinfo {author} {\bibfnamefont {A.~G.}\ \bibnamefont
  {Abanov}}\ and\ \bibinfo {author} {\bibfnamefont {D.~A.}\ \bibnamefont
  {Ivanov}},\ }\bibfield  {title} {\enquote {\bibinfo {title} {Allowed charge
  transfers between coherent conductors driven by a time-dependent
  scatterer},}\ }\href {\doibase 10.1103/PhysRevLett.100.086602} {\bibfield
  {journal} {\bibinfo  {journal} {Phys. Rev. Lett.}\ }\textbf {\bibinfo
  {volume} {100}},\ \bibinfo {pages} {086602} (\bibinfo {year}
  {2008})}\BibitemShut {NoStop}%
\bibitem [{\citenamefont {Abanov}\ and\ \citenamefont
  {Ivanov}(2009)}]{abanov:2009}%
  \BibitemOpen
  \bibfield  {author} {\bibinfo {author} {\bibfnamefont {A.~G.}\ \bibnamefont
  {Abanov}}\ and\ \bibinfo {author} {\bibfnamefont {D.~A.}\ \bibnamefont
  {Ivanov}},\ }\bibfield  {title} {\enquote {\bibinfo {title} {Factorization of
  quantum charge transport for noninteracting fermions},}\ }\href {\doibase
  10.1103/PhysRevB.79.205315} {\bibfield  {journal} {\bibinfo  {journal} {Phys.
  Rev. B}\ }\textbf {\bibinfo {volume} {79}},\ \bibinfo {pages} {205315}
  (\bibinfo {year} {2009})}\BibitemShut {NoStop}%
\bibitem [{\citenamefont {Kambly}\ and\ \citenamefont
  {Ivanov}(2009)}]{kambly:2009}%
  \BibitemOpen
  \bibfield  {author} {\bibinfo {author} {\bibfnamefont {D.}~\bibnamefont
  {Kambly}}\ and\ \bibinfo {author} {\bibfnamefont {D.~A.}\ \bibnamefont
  {Ivanov}},\ }\bibfield  {title} {\enquote {\bibinfo {title} {Statistics of
  quantum transfer of noninteracting fermions in multiterminal junctions},}\
  }\href {\doibase 10.1103/PhysRevB.80.193306} {\bibfield  {journal} {\bibinfo
  {journal} {Phys. Rev. B}\ }\textbf {\bibinfo {volume} {80}},\ \bibinfo
  {pages} {193306} (\bibinfo {year} {2009})}\BibitemShut {NoStop}%
\bibitem [{\citenamefont {Kambly}\ \emph {et~al.}(2011)\citenamefont {Kambly},
  \citenamefont {Flindt},\ and\ \citenamefont {B\"uttiker}}]{kambly:2011}%
  \BibitemOpen
  \bibfield  {author} {\bibinfo {author} {\bibfnamefont {D.}~\bibnamefont
  {Kambly}}, \bibinfo {author} {\bibfnamefont {C.}~\bibnamefont {Flindt}}, \
  and\ \bibinfo {author} {\bibfnamefont {M.}~\bibnamefont {B\"uttiker}},\
  }\bibfield  {title} {\enquote {\bibinfo {title} {Factorial cumulants reveal
  interactions in counting statistics},}\ }\href {\doibase
  10.1103/PhysRevB.83.075432} {\bibfield  {journal} {\bibinfo  {journal} {Phys.
  Rev. B}\ }\textbf {\bibinfo {volume} {83}},\ \bibinfo {pages} {075432}
  (\bibinfo {year} {2011})}\BibitemShut {NoStop}%
\bibitem [{\citenamefont {Klich}\ and\ \citenamefont
  {Levitov}(2009)}]{klich:2009}%
  \BibitemOpen
  \bibfield  {author} {\bibinfo {author} {\bibfnamefont {I.}~\bibnamefont
  {Klich}}\ and\ \bibinfo {author} {\bibfnamefont {L.~S.}\ \bibnamefont
  {Levitov}},\ }\bibfield  {title} {\enquote {\bibinfo {title} {Quantum noise
  as an entanglement meter},}\ }\href {\doibase 10.1103/PhysRevLett.102.100502}
  {\bibfield  {journal} {\bibinfo  {journal} {Phys. Rev. Lett.}\ }\textbf
  {\bibinfo {volume} {102}},\ \bibinfo {pages} {100502} (\bibinfo {year}
  {2009})}\BibitemShut {NoStop}%
\bibitem [{\citenamefont {Song}\ \emph {et~al.}(2011)\citenamefont {Song},
  \citenamefont {Flindt}, \citenamefont {Rachel}, \citenamefont {Klich},\ and\
  \citenamefont {Le~Hur}}]{song:2011}%
  \BibitemOpen
  \bibfield  {author} {\bibinfo {author} {\bibfnamefont {H.~F.}\ \bibnamefont
  {Song}}, \bibinfo {author} {\bibfnamefont {C.}~\bibnamefont {Flindt}},
  \bibinfo {author} {\bibfnamefont {S.}~\bibnamefont {Rachel}}, \bibinfo
  {author} {\bibfnamefont {I.}~\bibnamefont {Klich}}, \ and\ \bibinfo {author}
  {\bibfnamefont {K.}~\bibnamefont {Le~Hur}},\ }\bibfield  {title} {\enquote
  {\bibinfo {title} {Entanglement entropy from charge statistics: Exact
  relations for noninteracting many-body systems},}\ }\href {\doibase
  10.1103/PhysRevB.83.161408} {\bibfield  {journal} {\bibinfo  {journal} {Phys.
  Rev. B}\ }\textbf {\bibinfo {volume} {83}},\ \bibinfo {pages} {161408}
  (\bibinfo {year} {2011})}\BibitemShut {NoStop}%
\bibitem [{\citenamefont {Song}\ \emph {et~al.}(2012)\citenamefont {Song},
  \citenamefont {Rachel}, \citenamefont {Flindt}, \citenamefont {Klich},
  \citenamefont {Laflorencie},\ and\ \citenamefont {Le~Hur}}]{song:2012}%
  \BibitemOpen
  \bibfield  {author} {\bibinfo {author} {\bibfnamefont {H.~F.}\ \bibnamefont
  {Song}}, \bibinfo {author} {\bibfnamefont {S.}~\bibnamefont {Rachel}},
  \bibinfo {author} {\bibfnamefont {C.}~\bibnamefont {Flindt}}, \bibinfo
  {author} {\bibfnamefont {I.}~\bibnamefont {Klich}}, \bibinfo {author}
  {\bibfnamefont {N.}~\bibnamefont {Laflorencie}}, \ and\ \bibinfo {author}
  {\bibfnamefont {K.}~\bibnamefont {Le~Hur}},\ }\bibfield  {title} {\enquote
  {\bibinfo {title} {Bipartite fluctuations as a probe of many-body
  entanglement},}\ }\href {\doibase 10.1103/PhysRevB.85.035409} {\bibfield
  {journal} {\bibinfo  {journal} {Phys. Rev. B}\ }\textbf {\bibinfo {volume}
  {85}},\ \bibinfo {pages} {035409} (\bibinfo {year} {2012})}\BibitemShut
  {NoStop}%
\bibitem [{\citenamefont {Petrescu}\ \emph {et~al.}(2014)\citenamefont
  {Petrescu}, \citenamefont {Song}, \citenamefont {Rachel}, \citenamefont
  {Ristivojevic}, \citenamefont {Flindt}, \citenamefont {Laflorencie},
  \citenamefont {Klich}, \citenamefont {Regnault},\ and\ \citenamefont
  {Le~Hur}}]{petrescu:2014}%
  \BibitemOpen
  \bibfield  {author} {\bibinfo {author} {\bibfnamefont {A.}~\bibnamefont
  {Petrescu}}, \bibinfo {author} {\bibfnamefont {H.~F.}\ \bibnamefont {Song}},
  \bibinfo {author} {\bibfnamefont {S.}~\bibnamefont {Rachel}}, \bibinfo
  {author} {\bibfnamefont {Z.}~\bibnamefont {Ristivojevic}}, \bibinfo {author}
  {\bibfnamefont {C.}~\bibnamefont {Flindt}}, \bibinfo {author} {\bibfnamefont
  {N.}~\bibnamefont {Laflorencie}}, \bibinfo {author} {\bibfnamefont
  {I.}~\bibnamefont {Klich}}, \bibinfo {author} {\bibfnamefont
  {N.}~\bibnamefont {Regnault}}, \ and\ \bibinfo {author} {\bibfnamefont
  {K.}~\bibnamefont {Le~Hur}},\ }\bibfield  {title} {\enquote {\bibinfo {title}
  {Fluctuations and entanglement spectrum in quantum hall states},}\ }\href
  {\doibase 10.1088/1742-5468/2014/10/P10005} {\bibfield  {journal} {\bibinfo
  {journal} {J. Stat. Mech.}\ ,\ \bibinfo {pages} {P10005}} (\bibinfo {year}
  {2014})}\BibitemShut {NoStop}%
\bibitem [{\citenamefont {Thomas}\ and\ \citenamefont
  {Flindt}(2015)}]{thomas:2015}%
  \BibitemOpen
  \bibfield  {author} {\bibinfo {author} {\bibfnamefont {K.~H.}\ \bibnamefont
  {Thomas}}\ and\ \bibinfo {author} {\bibfnamefont {C.}~\bibnamefont
  {Flindt}},\ }\bibfield  {title} {\enquote {\bibinfo {title} {Entanglement
  entropy in dynamic quantum-coherent conductors},}\ }\href {\doibase
  10.1103/PhysRevB.91.125406} {\bibfield  {journal} {\bibinfo  {journal} {Phys.
  Rev. B}\ }\textbf {\bibinfo {volume} {91}},\ \bibinfo {pages} {125406}
  (\bibinfo {year} {2015})}\BibitemShut {NoStop}%
\bibitem [{\citenamefont {F\"orster}\ and\ \citenamefont
  {B\"uttiker}(2008)}]{forster:2008}%
  \BibitemOpen
  \bibfield  {author} {\bibinfo {author} {\bibfnamefont {H.}~\bibnamefont
  {F\"orster}}\ and\ \bibinfo {author} {\bibfnamefont {M.}~\bibnamefont
  {B\"uttiker}},\ }\bibfield  {title} {\enquote {\bibinfo {title} {Fluctuation
  relations without microreversibility in nonlinear transport},}\ }\href
  {\doibase 10.1103/PhysRevLett.101.136805} {\bibfield  {journal} {\bibinfo
  {journal} {Phys. Rev. Lett.}\ }\textbf {\bibinfo {volume} {101}},\ \bibinfo
  {pages} {136805} (\bibinfo {year} {2008})}\BibitemShut {NoStop}%
\bibitem [{\citenamefont {Saito}\ and\ \citenamefont
  {Utsumi}(2008)}]{saito:2008}%
  \BibitemOpen
  \bibfield  {author} {\bibinfo {author} {\bibfnamefont {K.}~\bibnamefont
  {Saito}}\ and\ \bibinfo {author} {\bibfnamefont {Y.}~\bibnamefont {Utsumi}},\
  }\bibfield  {title} {\enquote {\bibinfo {title} {Symmetry in full counting
  statistics, fluctuation theorem, and relations among nonlinear transport
  coefficients in the presence of a magnetic field},}\ }\href {\doibase
  10.1103/PhysRevB.78.115429} {\bibfield  {journal} {\bibinfo  {journal} {Phys.
  Rev. B}\ }\textbf {\bibinfo {volume} {78}},\ \bibinfo {pages} {115429}
  (\bibinfo {year} {2008})}\BibitemShut {NoStop}%
\bibitem [{\citenamefont {Esposito}\ \emph {et~al.}(2009)\citenamefont
  {Esposito}, \citenamefont {Harbola},\ and\ \citenamefont
  {Mukamel}}]{esposito:2009}%
  \BibitemOpen
  \bibfield  {author} {\bibinfo {author} {\bibfnamefont {M.}~\bibnamefont
  {Esposito}}, \bibinfo {author} {\bibfnamefont {U.}~\bibnamefont {Harbola}}, \
  and\ \bibinfo {author} {\bibfnamefont {S.}~\bibnamefont {Mukamel}},\
  }\bibfield  {title} {\enquote {\bibinfo {title} {Nonequilibrium fluctuations,
  fluctuation theorems, and counting statistics in quantum systems},}\ }\href
  {\doibase 10.1103/RevModPhys.81.1665} {\bibfield  {journal} {\bibinfo
  {journal} {Rev. Mod. Phys.}\ }\textbf {\bibinfo {volume} {81}},\ \bibinfo
  {pages} {1665} (\bibinfo {year} {2009})}\BibitemShut {NoStop}%
\bibitem [{\citenamefont {Brandes}(2008)}]{brandes:2008}%
  \BibitemOpen
  \bibfield  {author} {\bibinfo {author} {\bibfnamefont {T.}~\bibnamefont
  {Brandes}},\ }\bibfield  {title} {\enquote {\bibinfo {title} {Waiting times
  and noise in single particle transport},}\ }\href {\doibase
  10.1002/andp.200810306} {\bibfield  {journal} {\bibinfo  {journal} {Ann.
  Phys.}\ }\textbf {\bibinfo {volume} {17}},\ \bibinfo {pages} {477} (\bibinfo
  {year} {2008})}\BibitemShut {NoStop}%
\bibitem [{\citenamefont {Welack}\ \emph {et~al.}(2009)\citenamefont {Welack},
  \citenamefont {Mukamel},\ and\ \citenamefont {Yan}}]{welack:2009}%
  \BibitemOpen
  \bibfield  {author} {\bibinfo {author} {\bibfnamefont {S.}~\bibnamefont
  {Welack}}, \bibinfo {author} {\bibfnamefont {S.}~\bibnamefont {Mukamel}}, \
  and\ \bibinfo {author} {\bibfnamefont {Y.}~\bibnamefont {Yan}},\ }\bibfield
  {title} {\enquote {\bibinfo {title} {Waiting time distributions of electron
  transfers through quantum dot Aharonov-Bohm interferometers},}\ }\href
  {\doibase 10.1209/0295-5075/85/57008} {\bibfield  {journal} {\bibinfo
  {journal} {Europhys. Lett.}\ }\textbf {\bibinfo {volume} {85}},\ \bibinfo
  {pages} {57008} (\bibinfo {year} {2009})}\BibitemShut {NoStop}%
\bibitem [{\citenamefont {Albert}\ \emph {et~al.}(2011)\citenamefont {Albert},
  \citenamefont {Flindt},\ and\ \citenamefont {B\"uttiker}}]{albert:2011}%
  \BibitemOpen
  \bibfield  {author} {\bibinfo {author} {\bibfnamefont {M.}~\bibnamefont
  {Albert}}, \bibinfo {author} {\bibfnamefont {C.}~\bibnamefont {Flindt}}, \
  and\ \bibinfo {author} {\bibfnamefont {M.}~\bibnamefont {B\"uttiker}},\
  }\bibfield  {title} {\enquote {\bibinfo {title} {Distributions of waiting
  times of dynamic single-electron emitters},}\ }\href {\doibase
  10.1103/PhysRevLett.107.086805} {\bibfield  {journal} {\bibinfo  {journal}
  {Phys. Rev. Lett.}\ }\textbf {\bibinfo {volume} {107}},\ \bibinfo {pages}
  {086805} (\bibinfo {year} {2011})}\BibitemShut {NoStop}%
\bibitem [{\citenamefont {Rajabi}\ \emph {et~al.}(2013)\citenamefont {Rajabi},
  \citenamefont {P\"oltl},\ and\ \citenamefont {Governale}}]{rajabi:2013}%
  \BibitemOpen
  \bibfield  {author} {\bibinfo {author} {\bibfnamefont {L.}~\bibnamefont
  {Rajabi}}, \bibinfo {author} {\bibfnamefont {C.}~\bibnamefont {P\"oltl}}, \
  and\ \bibinfo {author} {\bibfnamefont {M.}~\bibnamefont {Governale}},\
  }\bibfield  {title} {\enquote {\bibinfo {title} {Waiting time distributions
  for the transport through a quantum-dot tunnel coupled to one normal and one
  superconducting lead},}\ }\href {\doibase 10.1103/PhysRevLett.111.067002}
  {\bibfield  {journal} {\bibinfo  {journal} {Phys. Rev. Lett.}\ }\textbf
  {\bibinfo {volume} {111}},\ \bibinfo {pages} {067002} (\bibinfo {year}
  {2013})}\BibitemShut {NoStop}%
\bibitem [{\citenamefont {Sothmann}(2014)}]{sothman:2014}%
  \BibitemOpen
  \bibfield  {author} {\bibinfo {author} {\bibfnamefont {B.}~\bibnamefont
  {Sothmann}},\ }\bibfield  {title} {\enquote {\bibinfo {title} {Electronic
  waiting-time distribution of a quantum-dot spin valve},}\ }\href {\doibase
  10.1103/PhysRevB.90.155315} {\bibfield  {journal} {\bibinfo  {journal} {Phys.
  Rev. B}\ }\textbf {\bibinfo {volume} {90}},\ \bibinfo {pages} {155315}
  (\bibinfo {year} {2014})}\BibitemShut {NoStop}%
\bibitem [{\citenamefont {Thomas}\ and\ \citenamefont
  {Flindt}(2013)}]{thomas:2013}%
  \BibitemOpen
  \bibfield  {author} {\bibinfo {author} {\bibfnamefont {K.~H.}\ \bibnamefont
  {Thomas}}\ and\ \bibinfo {author} {\bibfnamefont {C.}~\bibnamefont
  {Flindt}},\ }\bibfield  {title} {\enquote {\bibinfo {title} {Electron waiting
  times in non-markovian quantum transport},}\ }\href {\doibase
  10.1103/PhysRevB.87.121405} {\bibfield  {journal} {\bibinfo  {journal} {Phys.
  Rev. B}\ }\textbf {\bibinfo {volume} {87}},\ \bibinfo {pages} {121405}
  (\bibinfo {year} {2013})}\BibitemShut {NoStop}%
\bibitem [{\citenamefont {Albert}\ \emph {et~al.}(2012)\citenamefont {Albert},
  \citenamefont {Haack}, \citenamefont {Flindt},\ and\ \citenamefont
  {B\"uttiker}}]{albert:2012}%
  \BibitemOpen
  \bibfield  {author} {\bibinfo {author} {\bibfnamefont {M.}~\bibnamefont
  {Albert}}, \bibinfo {author} {\bibfnamefont {G.}~\bibnamefont {Haack}},
  \bibinfo {author} {\bibfnamefont {C.}~\bibnamefont {Flindt}}, \ and\ \bibinfo
  {author} {\bibfnamefont {M.}~\bibnamefont {B\"uttiker}},\ }\bibfield  {title}
  {\enquote {\bibinfo {title} {Electron waiting times in mesoscopic
  conductors},}\ }\href {\doibase 10.1103/PhysRevLett.108.186806} {\bibfield
  {journal} {\bibinfo  {journal} {Phys. Rev. Lett.}\ }\textbf {\bibinfo
  {volume} {108}},\ \bibinfo {pages} {186806} (\bibinfo {year}
  {2012})}\BibitemShut {NoStop}%
\bibitem [{\citenamefont {Dasenbrook}\ \emph {et~al.}(2014)\citenamefont
  {Dasenbrook}, \citenamefont {Flindt},\ and\ \citenamefont
  {B\"uttiker}}]{dasenbrook:2014}%
  \BibitemOpen
  \bibfield  {author} {\bibinfo {author} {\bibfnamefont {D.}~\bibnamefont
  {Dasenbrook}}, \bibinfo {author} {\bibfnamefont {C.}~\bibnamefont {Flindt}},
  \ and\ \bibinfo {author} {\bibfnamefont {M.}~\bibnamefont {B\"uttiker}},\
  }\bibfield  {title} {\enquote {\bibinfo {title} {Floquet theory of electron
  waiting times in quantum-coherent conductors},}\ }\href {\doibase
  10.1103/PhysRevLett.112.146801} {\bibfield  {journal} {\bibinfo  {journal}
  {Phys. Rev. Lett.}\ }\textbf {\bibinfo {volume} {112}},\ \bibinfo {pages}
  {146801} (\bibinfo {year} {2014})}\BibitemShut {NoStop}%
\bibitem [{\citenamefont {Haack}\ \emph {et~al.}(2014)\citenamefont {Haack},
  \citenamefont {Albert},\ and\ \citenamefont {Flindt}}]{haack:2014}%
  \BibitemOpen
  \bibfield  {author} {\bibinfo {author} {\bibfnamefont {G.}~\bibnamefont
  {Haack}}, \bibinfo {author} {\bibfnamefont {M.}~\bibnamefont {Albert}}, \
  and\ \bibinfo {author} {\bibfnamefont {C.}~\bibnamefont {Flindt}},\
  }\bibfield  {title} {\enquote {\bibinfo {title} {Distributions of electron
  waiting times in quantum-coherent conductors},}\ }\href {\doibase
  10.1103/PhysRevB.90.205429} {\bibfield  {journal} {\bibinfo  {journal} {Phys.
  Rev. B}\ }\textbf {\bibinfo {volume} {90}},\ \bibinfo {pages} {205429}
  (\bibinfo {year} {2014})}\BibitemShut {NoStop}%
\bibitem [{\citenamefont {Thomas}\ and\ \citenamefont
  {Flindt}(2014)}]{thomas:2014}%
  \BibitemOpen
  \bibfield  {author} {\bibinfo {author} {\bibfnamefont {K.~H.}\ \bibnamefont
  {Thomas}}\ and\ \bibinfo {author} {\bibfnamefont {C.}~\bibnamefont
  {Flindt}},\ }\bibfield  {title} {\enquote {\bibinfo {title} {Waiting time
  distributions of noninteracting fermions on a tight-binding chain},}\ }\href
  {\doibase 10.1103/PhysRevB.89.245420} {\bibfield  {journal} {\bibinfo
  {journal} {Phys. Rev. B}\ }\textbf {\bibinfo {volume} {89}},\ \bibinfo
  {pages} {245420} (\bibinfo {year} {2014})}\BibitemShut {NoStop}%
\bibitem [{\citenamefont {F\`eve}\ \emph {et~al.}(2007)\citenamefont {F\`eve},
  \citenamefont {Mah\'e}, \citenamefont {Berroir}, \citenamefont {Kontos},
  \citenamefont {Pla\ifmmode~\mbox{\c{c}}\else \c{c}\fi{}ais}, \citenamefont
  {Glattli}, \citenamefont {Cavanna}, \citenamefont {Etienne},\ and\
  \citenamefont {Jin}}]{feve:2007}%
  \BibitemOpen
  \bibfield  {author} {\bibinfo {author} {\bibfnamefont {G.}~\bibnamefont
  {F\`eve}}, \bibinfo {author} {\bibfnamefont {A.}~\bibnamefont {Mah\'e}},
  \bibinfo {author} {\bibfnamefont {J.-M.}\ \bibnamefont {Berroir}}, \bibinfo
  {author} {\bibfnamefont {T.}~\bibnamefont {Kontos}}, \bibinfo {author}
  {\bibfnamefont {B.}~\bibnamefont {Pla\ifmmode~\mbox{\c{c}}\else
  \c{c}\fi{}ais}}, \bibinfo {author} {\bibfnamefont {D.~C.}\ \bibnamefont
  {Glattli}}, \bibinfo {author} {\bibfnamefont {A.}~\bibnamefont {Cavanna}},
  \bibinfo {author} {\bibfnamefont {B.}~\bibnamefont {Etienne}}, \ and\
  \bibinfo {author} {\bibfnamefont {Y.}~\bibnamefont {Jin}},\ }\bibfield
  {title} {\enquote {\bibinfo {title} {An on-demand coherent single-electron
  source},}\ }\href {http://www.sciencemag.org/content/316/5828/1169.abstract}
  {\bibfield  {journal} {\bibinfo  {journal} {Science}\ }\textbf {\bibinfo
  {volume} {316}},\ \bibinfo {pages} {1169} (\bibinfo {year}
  {2007})}\BibitemShut {NoStop}%
\bibitem [{\citenamefont {Dubois}\ \emph
  {et~al.}(2013{\natexlab{a}})\citenamefont {Dubois}, \citenamefont {Jullien},
  \citenamefont {Portier}, \citenamefont {Roche}, \citenamefont {Cavanna},
  \citenamefont {Jin}, \citenamefont {Wegscheider}, \citenamefont {Roulleau},\
  and\ \citenamefont {Glattli}}]{dubois:2013}%
  \BibitemOpen
  \bibfield  {author} {\bibinfo {author} {\bibfnamefont {J.}~\bibnamefont
  {Dubois}}, \bibinfo {author} {\bibfnamefont {T.}~\bibnamefont {Jullien}},
  \bibinfo {author} {\bibfnamefont {F.}~\bibnamefont {Portier}}, \bibinfo
  {author} {\bibfnamefont {P.}~\bibnamefont {Roche}}, \bibinfo {author}
  {\bibfnamefont {A.}~\bibnamefont {Cavanna}}, \bibinfo {author} {\bibfnamefont
  {Y.}~\bibnamefont {Jin}}, \bibinfo {author} {\bibfnamefont {W.}~\bibnamefont
  {Wegscheider}}, \bibinfo {author} {\bibfnamefont {P.}~\bibnamefont
  {Roulleau}}, \ and\ \bibinfo {author} {\bibfnamefont {D.~C.}\ \bibnamefont
  {Glattli}},\ }\bibfield  {title} {\enquote {\bibinfo {title}
  {Minimal-excitation states for electron quantum optics using levitons},}\
  }\href {http://dx.doi.org/10.1038/nature12713} {\bibfield  {journal}
  {\bibinfo  {journal} {Nature}\ }\textbf {\bibinfo {volume} {502}},\ \bibinfo
  {pages} {659} (\bibinfo {year} {2013}{\natexlab{a}})}\BibitemShut {NoStop}%
\bibitem [{\citenamefont {Jullien}\ \emph {et~al.}(2014)\citenamefont
  {Jullien}, \citenamefont {Roulleau}, \citenamefont {Roche}, \citenamefont
  {Cavanna}, \citenamefont {Jin},\ and\ \citenamefont
  {Glattli}}]{jullien:2014}%
  \BibitemOpen
  \bibfield  {author} {\bibinfo {author} {\bibfnamefont {T.}~\bibnamefont
  {Jullien}}, \bibinfo {author} {\bibfnamefont {P.}~\bibnamefont {Roulleau}},
  \bibinfo {author} {\bibfnamefont {B.}~\bibnamefont {Roche}}, \bibinfo
  {author} {\bibfnamefont {A.}~\bibnamefont {Cavanna}}, \bibinfo {author}
  {\bibfnamefont {Y.}~\bibnamefont {Jin}}, \ and\ \bibinfo {author}
  {\bibfnamefont {D.~C.}\ \bibnamefont {Glattli}},\ }\bibfield  {title}
  {\enquote {\bibinfo {title} {Quantum tomography of an electron},}\ }\href
  {\doibase 10.1038/nature13821} {\bibfield  {journal} {\bibinfo  {journal}
  {Nature}\ }\textbf {\bibinfo {volume} {514}},\ \bibinfo {pages} {603}
  (\bibinfo {year} {2014})}\BibitemShut {NoStop}%
\bibitem [{\citenamefont {Albert}\ and\ \citenamefont
  {Devillard}(2014)}]{albert:2014}%
  \BibitemOpen
  \bibfield  {author} {\bibinfo {author} {\bibfnamefont {M.}~\bibnamefont
  {Albert}}\ and\ \bibinfo {author} {\bibfnamefont {P.}~\bibnamefont
  {Devillard}},\ }\bibfield  {title} {\enquote {\bibinfo {title} {Waiting time
  distribution for trains of quantized electron pulses},}\ }\href {\doibase
  10.1103/PhysRevB.90.035431} {\bibfield  {journal} {\bibinfo  {journal} {Phys.
  Rev. B}\ }\textbf {\bibinfo {volume} {90}},\ \bibinfo {pages} {035431}
  (\bibinfo {year} {2014})}\BibitemShut {NoStop}%
\bibitem [{\citenamefont {Fletcher}\ \emph {et~al.}(2013)\citenamefont
  {Fletcher}, \citenamefont {See}, \citenamefont {Howe}, \citenamefont
  {Pepper}, \citenamefont {Giblin}, \citenamefont {Griffiths}, \citenamefont
  {Jones}, \citenamefont {Farrer}, \citenamefont {Ritchie}, \citenamefont
  {Janssen},\ and\ \citenamefont {Kataoka}}]{fletcher:2013}%
  \BibitemOpen
  \bibfield  {author} {\bibinfo {author} {\bibfnamefont {J.~D.}\ \bibnamefont
  {Fletcher}}, \bibinfo {author} {\bibfnamefont {P.}~\bibnamefont {See}},
  \bibinfo {author} {\bibfnamefont {H.}~\bibnamefont {Howe}}, \bibinfo {author}
  {\bibfnamefont {M.}~\bibnamefont {Pepper}}, \bibinfo {author} {\bibfnamefont
  {S.~P.}\ \bibnamefont {Giblin}}, \bibinfo {author} {\bibfnamefont {J.~P.}\
  \bibnamefont {Griffiths}}, \bibinfo {author} {\bibfnamefont {G.~A.~C.}\
  \bibnamefont {Jones}}, \bibinfo {author} {\bibfnamefont {I.}~\bibnamefont
  {Farrer}}, \bibinfo {author} {\bibfnamefont {D.~A.}\ \bibnamefont {Ritchie}},
  \bibinfo {author} {\bibfnamefont {T.~J. B.~M.}\ \bibnamefont {Janssen}}, \
  and\ \bibinfo {author} {\bibfnamefont {M.}~\bibnamefont {Kataoka}},\
  }\bibfield  {title} {\enquote {\bibinfo {title} {Clock-controlled emission of
  single-electron wave packets in a solid-state circuit},}\ }\href {\doibase
  10.1103/PhysRevLett.111.216807} {\bibfield  {journal} {\bibinfo  {journal}
  {Phys. Rev. Lett.}\ }\textbf {\bibinfo {volume} {111}},\ \bibinfo {pages}
  {216807} (\bibinfo {year} {2013})}\BibitemShut {NoStop}%
\bibitem [{\citenamefont {Thalineau}\ \emph {et~al.}(2014)\citenamefont
  {Thalineau}, \citenamefont {Wieck}, \citenamefont {B\"auerle},\ and\
  \citenamefont {Meunier}}]{thalineau:2014}%
  \BibitemOpen
  \bibfield  {author} {\bibinfo {author} {\bibfnamefont {R.}~\bibnamefont
  {Thalineau}}, \bibinfo {author} {\bibfnamefont {A.~D.}\ \bibnamefont
  {Wieck}}, \bibinfo {author} {\bibfnamefont {C.}~\bibnamefont {B\"auerle}}, \
  and\ \bibinfo {author} {\bibfnamefont {T.}~\bibnamefont {Meunier}},\
  }\href@noop {} {\enquote {\bibinfo {title} {Using a two-electron spin qubit
  to detect electrons flying above the fermi sea},}\ } (\bibinfo {year}
  {2014}),\ \bibinfo {note} {arXiv:1403.7770}\BibitemShut {NoStop}%
\bibitem [{\citenamefont {Cox}(1962)}]{cox:1962}%
  \BibitemOpen
  \bibfield  {author} {\bibinfo {author} {\bibfnamefont {D.~R.}\ \bibnamefont
  {Cox}},\ }\href@noop {} {\emph {\bibinfo {title} {Renewal Theory}}}\
  (\bibinfo  {publisher} {Chapman and Hall},\ \bibinfo {year}
  {1962})\BibitemShut {NoStop}%
\bibitem [{\citenamefont {Herman}\ \emph {et~al.}(2007)\citenamefont {Herman},
  \citenamefont {Ong}, \citenamefont {Usaj}, \citenamefont {Mathur},\ and\
  \citenamefont {Baranger}}]{herman:2007}%
  \BibitemOpen
  \bibfield  {author} {\bibinfo {author} {\bibfnamefont {D.}~\bibnamefont
  {Herman}}, \bibinfo {author} {\bibfnamefont {T.~T.}\ \bibnamefont {Ong}},
  \bibinfo {author} {\bibfnamefont {G.}~\bibnamefont {Usaj}}, \bibinfo {author}
  {\bibfnamefont {H.}~\bibnamefont {Mathur}}, \ and\ \bibinfo {author}
  {\bibfnamefont {H.~U.}\ \bibnamefont {Baranger}},\ }\bibfield  {title}
  {\enquote {\bibinfo {title} {Level spacings in random matrix theory and
  coulomb blockade peaks in quantum dots},}\ }\href {\doibase
  10.1103/PhysRevB.76.195448} {\bibfield  {journal} {\bibinfo  {journal} {Phys.
  Rev. B}\ }\textbf {\bibinfo {volume} {76}},\ \bibinfo {pages} {195448}
  (\bibinfo {year} {2007})}\BibitemShut {NoStop}%
\bibitem [{\citenamefont {Ivanov}\ \emph {et~al.}(1997)\citenamefont {Ivanov},
  \citenamefont {Lee},\ and\ \citenamefont {Levitov}}]{ivanov:1997}%
  \BibitemOpen
  \bibfield  {author} {\bibinfo {author} {\bibfnamefont {D.~A.}\ \bibnamefont
  {Ivanov}}, \bibinfo {author} {\bibfnamefont {H.~W.}\ \bibnamefont {Lee}}, \
  and\ \bibinfo {author} {\bibfnamefont {L.~S.}\ \bibnamefont {Levitov}},\
  }\bibfield  {title} {\enquote {\bibinfo {title} {Coherent states of
  alternating current},}\ }\href {\doibase 10.1103/PhysRevB.56.6839} {\bibfield
   {journal} {\bibinfo  {journal} {Phys. Rev. B}\ }\textbf {\bibinfo {volume}
  {56}},\ \bibinfo {pages} {6839} (\bibinfo {year} {1997})}\BibitemShut
  {NoStop}%
\bibitem [{\citenamefont {Keeling}\ \emph {et~al.}(2006)\citenamefont
  {Keeling}, \citenamefont {Klich},\ and\ \citenamefont
  {Levitov}}]{keeling:2006}%
  \BibitemOpen
  \bibfield  {author} {\bibinfo {author} {\bibfnamefont {J.}~\bibnamefont
  {Keeling}}, \bibinfo {author} {\bibfnamefont {I.}~\bibnamefont {Klich}}, \
  and\ \bibinfo {author} {\bibfnamefont {L.~S.}\ \bibnamefont {Levitov}},\
  }\bibfield  {title} {\enquote {\bibinfo {title} {Minimal excitation states of
  electrons in one-dimensional wires},}\ }\href
  {http://link.aps.org/doi/10.1103/PhysRevLett.97.116403} {\bibfield  {journal}
  {\bibinfo  {journal} {Phys. Rev. Lett.}\ }\textbf {\bibinfo {volume} {97}},\
  \bibinfo {pages} {116403} (\bibinfo {year} {2006})}\BibitemShut {NoStop}%
\bibitem [{\citenamefont {Brantut}\ \emph {et~al.}(2012)\citenamefont
  {Brantut}, \citenamefont {Meineke}, \citenamefont {Stadler}, \citenamefont
  {Krinner},\ and\ \citenamefont {Esslinger}}]{brantut:2012}%
  \BibitemOpen
  \bibfield  {author} {\bibinfo {author} {\bibfnamefont {J.-P.}\ \bibnamefont
  {Brantut}}, \bibinfo {author} {\bibfnamefont {J.}~\bibnamefont {Meineke}},
  \bibinfo {author} {\bibfnamefont {D.}~\bibnamefont {Stadler}}, \bibinfo
  {author} {\bibfnamefont {S.}~\bibnamefont {Krinner}}, \ and\ \bibinfo
  {author} {\bibfnamefont {T.}~\bibnamefont {Esslinger}},\ }\bibfield  {title}
  {\enquote {\bibinfo {title} {Conduction of ultracold fermions through a
  mesoscopic channel},}\ }\href {\doibase 10.1126/science.1223175} {\bibfield
  {journal} {\bibinfo  {journal} {Science}\ }\textbf {\bibinfo {volume}
  {337}},\ \bibinfo {pages} {1069} (\bibinfo {year} {2012})}\BibitemShut
  {NoStop}%
\bibitem [{\citenamefont {Vyas}\ and\ \citenamefont {Singh}(1988)}]{vyas:1988}%
  \BibitemOpen
  \bibfield  {author} {\bibinfo {author} {\bibfnamefont {R.}~\bibnamefont
  {Vyas}}\ and\ \bibinfo {author} {\bibfnamefont {S.}~\bibnamefont {Singh}},\
  }\bibfield  {title} {\enquote {\bibinfo {title} {Waiting-time distributions
  in the photodetection of squeezed light},}\ }\href {\doibase
  10.1103/PhysRevA.38.2423} {\bibfield  {journal} {\bibinfo  {journal} {Phys.
  Rev. A}\ }\textbf {\bibinfo {volume} {38}},\ \bibinfo {pages} {2423}
  (\bibinfo {year} {1988})}\BibitemShut {NoStop}%
\bibitem [{\citenamefont {Saito}\ \emph {et~al.}(1992)\citenamefont {Saito},
  \citenamefont {Endo}, \citenamefont {Kodama}, \citenamefont {Tonomura},
  \citenamefont {Fukuhara},\ and\ \citenamefont {Ohbayashi}}]{saito:1992}%
  \BibitemOpen
  \bibfield  {author} {\bibinfo {author} {\bibfnamefont {S.}~\bibnamefont
  {Saito}}, \bibinfo {author} {\bibfnamefont {J.}~\bibnamefont {Endo}},
  \bibinfo {author} {\bibfnamefont {T.}~\bibnamefont {Kodama}}, \bibinfo
  {author} {\bibfnamefont {A.}~\bibnamefont {Tonomura}}, \bibinfo {author}
  {\bibfnamefont {A.}~\bibnamefont {Fukuhara}}, \ and\ \bibinfo {author}
  {\bibfnamefont {K.}~\bibnamefont {Ohbayashi}},\ }\bibfield  {title} {\enquote
  {\bibinfo {title} {Electron counting theory},}\ }\href {\doibase
  10.1016/0375-9601(92)90003-5} {\bibfield  {journal} {\bibinfo  {journal}
  {Phys. Lett. A}\ }\textbf {\bibinfo {volume} {162}},\ \bibinfo {pages} {442}
  (\bibinfo {year} {1992})}\BibitemShut {NoStop}%
\bibitem [{\citenamefont {Hassler}\ \emph {et~al.}(2008)\citenamefont
  {Hassler}, \citenamefont {Suslov}, \citenamefont {Graf}, \citenamefont
  {Lebedev}, \citenamefont {Lesovik},\ and\ \citenamefont
  {Blatter}}]{hassler:2008}%
  \BibitemOpen
  \bibfield  {author} {\bibinfo {author} {\bibfnamefont {F.}~\bibnamefont
  {Hassler}}, \bibinfo {author} {\bibfnamefont {M.~V.}\ \bibnamefont {Suslov}},
  \bibinfo {author} {\bibfnamefont {G.~M.}\ \bibnamefont {Graf}}, \bibinfo
  {author} {\bibfnamefont {M.~V.}\ \bibnamefont {Lebedev}}, \bibinfo {author}
  {\bibfnamefont {G.~B.}\ \bibnamefont {Lesovik}}, \ and\ \bibinfo {author}
  {\bibfnamefont {G.}~\bibnamefont {Blatter}},\ }\bibfield  {title} {\enquote
  {\bibinfo {title} {Wave-packet formalism of full counting statistics},}\
  }\href {\doibase 10.1103/PhysRevB.78.165330} {\bibfield  {journal} {\bibinfo
  {journal} {Phys. Rev. B}\ }\textbf {\bibinfo {volume} {78}},\ \bibinfo
  {pages} {165330} (\bibinfo {year} {2008})}\BibitemShut {NoStop}%
\bibitem [{\citenamefont {Giuliani}\ and\ \citenamefont
  {Vignale}(2005)}]{giuliani:2005}%
  \BibitemOpen
  \bibfield  {author} {\bibinfo {author} {\bibfnamefont {G.}~\bibnamefont
  {Giuliani}}\ and\ \bibinfo {author} {\bibfnamefont {G.}~\bibnamefont
  {Vignale}},\ }\href@noop {} {\emph {\bibinfo {title} {Quantum Theory of the
  Electron Liquid}}}\ (\bibinfo  {publisher} {Cambridge University Press},\
  \bibinfo {year} {2005})\BibitemShut {NoStop}%
\bibitem [{\citenamefont {Tang}\ \emph {et~al.}(2014)\citenamefont {Tang},
  \citenamefont {Xu},\ and\ \citenamefont {Wang}}]{tang:2014}%
  \BibitemOpen
  \bibfield  {author} {\bibinfo {author} {\bibfnamefont {G.-M.}\ \bibnamefont
  {Tang}}, \bibinfo {author} {\bibfnamefont {F.}~\bibnamefont {Xu}}, \ and\
  \bibinfo {author} {\bibfnamefont {J.}~\bibnamefont {Wang}},\ }\bibfield
  {title} {\enquote {\bibinfo {title} {Waiting time distribution of quantum
  electronic transport in the transient regime},}\ }\href {\doibase
  10.1103/PhysRevB.89.205310} {\bibfield  {journal} {\bibinfo  {journal} {Phys.
  Rev. B}\ }\textbf {\bibinfo {volume} {89}},\ \bibinfo {pages} {205310}
  (\bibinfo {year} {2014})}\BibitemShut {NoStop}%
\bibitem [{\citenamefont {Pedersen}\ and\ \citenamefont
  {B\"uttiker}(1998)}]{pedersen:1998}%
  \BibitemOpen
  \bibfield  {author} {\bibinfo {author} {\bibfnamefont {M.~H.}\ \bibnamefont
  {Pedersen}}\ and\ \bibinfo {author} {\bibfnamefont {M.}~\bibnamefont
  {B\"uttiker}},\ }\bibfield  {title} {\enquote {\bibinfo {title} {Scattering
  theory of photon-assisted electron transport},}\ }\href {\doibase
  10.1103/PhysRevB.58.12993} {\bibfield  {journal} {\bibinfo  {journal} {Phys.
  Rev. B}\ }\textbf {\bibinfo {volume} {58}},\ \bibinfo {pages} {12993}
  (\bibinfo {year} {1998})}\BibitemShut {NoStop}%
\bibitem [{\citenamefont {Moskalets}\ and\ \citenamefont
  {B\"uttiker}(2002)}]{moskalets:2002}%
  \BibitemOpen
  \bibfield  {author} {\bibinfo {author} {\bibfnamefont {M.}~\bibnamefont
  {Moskalets}}\ and\ \bibinfo {author} {\bibfnamefont {M.}~\bibnamefont
  {B\"uttiker}},\ }\bibfield  {title} {\enquote {\bibinfo {title} {Floquet
  scattering theory of quantum pumps},}\ }\href
  {http://link.aps.org/doi/10.1103/PhysRevB.66.205320} {\bibfield  {journal}
  {\bibinfo  {journal} {Phys. Rev. B}\ }\textbf {\bibinfo {volume} {66}},\
  \bibinfo {pages} {205320} (\bibinfo {year} {2002})}\BibitemShut {NoStop}%
\bibitem [{\citenamefont {Shelankov}\ and\ \citenamefont
  {Rammer}(2008)}]{shelankov:2008}%
  \BibitemOpen
  \bibfield  {author} {\bibinfo {author} {\bibfnamefont {A.}~\bibnamefont
  {Shelankov}}\ and\ \bibinfo {author} {\bibfnamefont {J.}~\bibnamefont
  {Rammer}},\ }\bibfield  {title} {\enquote {\bibinfo {title} {Counting
  statistics of interfering Bose-Einstein condensates},}\ }\href {\doibase
  10.1209/0295-5075/83/60002} {\bibfield  {journal} {\bibinfo  {journal} {Europhys. Lett.}\
  }\textbf {\bibinfo {volume} {83}},\ \bibinfo {pages} {60002} (\bibinfo {year}
  {2008})}\BibitemShut {NoStop}%
\bibitem [{\citenamefont {Rammer}\ and\ \citenamefont
  {Shelankov}(2012)}]{rammer:2012}%
  \BibitemOpen
  \bibfield  {author} {\bibinfo {author} {\bibfnamefont {J.}~\bibnamefont
  {Rammer}}\ and\ \bibinfo {author} {\bibfnamefont {A.}~\bibnamefont
  {Shelankov}},\ }\bibfield  {title} {\enquote {\bibinfo {title} {Counting
  quantum fluctuations of particle density},}\ }\href {\doibase
  10.1002/andp.201100277} {\bibfield  {journal} {\bibinfo  {journal} {Ann.
  Phys.}\ }\textbf {\bibinfo {volume} {524}},\ \bibinfo {pages} {163--174}
  (\bibinfo {year} {2012})}\BibitemShut {NoStop}%
\bibitem [{\citenamefont {Liu}\ \emph {et~al.}(1998)\citenamefont {Liu},
  \citenamefont {Odom}, \citenamefont {Yamamoto},\ and\ \citenamefont
  {Tarucha}}]{liu:1998}%
  \BibitemOpen
  \bibfield  {author} {\bibinfo {author} {\bibfnamefont {R.~C.}\ \bibnamefont
  {Liu}}, \bibinfo {author} {\bibfnamefont {B.}~\bibnamefont {Odom}}, \bibinfo
  {author} {\bibfnamefont {Y.}~\bibnamefont {Yamamoto}}, \ and\ \bibinfo
  {author} {\bibfnamefont {S.}~\bibnamefont {Tarucha}},\ }\bibfield  {title}
  {\enquote {\bibinfo {title} {Quantum interference in electron collision},}\
  }\href {\doibase 10.1038/34611} {\bibfield  {journal} {\bibinfo  {journal}
  {Nature}\ }\textbf {\bibinfo {volume} {391}},\ \bibinfo {pages} {263--265}
  (\bibinfo {year} {1998})}\BibitemShut {NoStop}%
\bibitem [{\citenamefont {Burkard}\ \emph {et~al.}(2000)\citenamefont
  {Burkard}, \citenamefont {Loss},\ and\ \citenamefont
  {Sukhorukov}}]{burkard:2000}%
  \BibitemOpen
  \bibfield  {author} {\bibinfo {author} {\bibfnamefont {G.}~\bibnamefont
  {Burkard}}, \bibinfo {author} {\bibfnamefont {D.}~\bibnamefont {Loss}}, \
  and\ \bibinfo {author} {\bibfnamefont {E.~V.}\ \bibnamefont {Sukhorukov}},\
  }\bibfield  {title} {\enquote {\bibinfo {title} {Noise of entangled
  electrons: Bunching and antibunching},}\ }\href {\doibase
  10.1103/PhysRevB.61.R16303} {\bibfield  {journal} {\bibinfo  {journal} {Phys.
  Rev. B}\ }\textbf {\bibinfo {volume} {61}},\ \bibinfo {pages} {R16303}
  (\bibinfo {year} {2000})}\BibitemShut {NoStop}%
\bibitem [{\citenamefont {Ol'khovskaya}\ \emph {et~al.}(2008)\citenamefont
  {Ol'khovskaya}, \citenamefont {Splettstoesser}, \citenamefont {Moskalets},\
  and\ \citenamefont {B\"uttiker}}]{olkhovskaya:2008}%
  \BibitemOpen
  \bibfield  {author} {\bibinfo {author} {\bibfnamefont {S.}~\bibnamefont
  {Ol'khovskaya}}, \bibinfo {author} {\bibfnamefont {J.}~\bibnamefont
  {Splettstoesser}}, \bibinfo {author} {\bibfnamefont {M.}~\bibnamefont
  {Moskalets}}, \ and\ \bibinfo {author} {\bibfnamefont {M.}~\bibnamefont
  {B\"uttiker}},\ }\bibfield  {title} {\enquote {\bibinfo {title} {Shot noise
  of a mesoscopic two-particle collider},}\ }\href
  {http://link.aps.org/doi/10.1103/PhysRevLett.101.166802} {\bibfield
  {journal} {\bibinfo  {journal} {Phys. Rev. Lett.}\ }\textbf {\bibinfo
  {volume} {101}},\ \bibinfo {pages} {166802} (\bibinfo {year}
  {2008})}\BibitemShut {NoStop}%
\bibitem [{\citenamefont {Giovannetti}\ \emph {et~al.}(2006)\citenamefont
  {Giovannetti}, \citenamefont {Frustaglia}, \citenamefont {Taddei},\ and\
  \citenamefont {Fazio}}]{giovannetti:2006}%
  \BibitemOpen
  \bibfield  {author} {\bibinfo {author} {\bibfnamefont {V.}~\bibnamefont
  {Giovannetti}}, \bibinfo {author} {\bibfnamefont {D.}~\bibnamefont
  {Frustaglia}}, \bibinfo {author} {\bibfnamefont {F.}~\bibnamefont {Taddei}},
  \ and\ \bibinfo {author} {\bibfnamefont {R.}~\bibnamefont {Fazio}},\
  }\bibfield  {title} {\enquote {\bibinfo {title} {Electronic Hong-Ou-Mandel
  interferometer for multimode entanglement detection},}\ }\href {\doibase
  10.1103/PhysRevB.74.115315} {\bibfield  {journal} {\bibinfo  {journal} {Phys.
  Rev. B}\ }\textbf {\bibinfo {volume} {74}},\ \bibinfo {pages} {115315}
  (\bibinfo {year} {2006})}\BibitemShut {NoStop}%
\bibitem [{\citenamefont {Jonckheere}\ \emph {et~al.}(2012)\citenamefont
  {Jonckheere}, \citenamefont {Rech}, \citenamefont {Wahl},\ and\ \citenamefont
  {Martin}}]{jonckheere:2012}%
  \BibitemOpen
  \bibfield  {author} {\bibinfo {author} {\bibfnamefont {T.}~\bibnamefont
  {Jonckheere}}, \bibinfo {author} {\bibfnamefont {J.}~\bibnamefont {Rech}},
  \bibinfo {author} {\bibfnamefont {C.}~\bibnamefont {Wahl}}, \ and\ \bibinfo
  {author} {\bibfnamefont {T.}~\bibnamefont {Martin}},\ }\bibfield  {title}
  {\enquote {\bibinfo {title} {Electron and hole Hong-Ou-Mandel
  interferometry},}\ }\href {\doibase 10.1103/PhysRevB.86.125425} {\bibfield
  {journal} {\bibinfo  {journal} {Phys. Rev. B}\ }\textbf {\bibinfo {volume}
  {86}},\ \bibinfo {pages} {125425} (\bibinfo {year} {2012})}\BibitemShut
  {NoStop}%
\bibitem [{\citenamefont {Bocquillon}\ \emph {et~al.}(2013)\citenamefont
  {Bocquillon}, \citenamefont {Freulon}, \citenamefont {Berroir}, \citenamefont
  {Degiovanni}, \citenamefont {Pla\c{c}ais}, \citenamefont {Cavanna},
  \citenamefont {Jin},\ and\ \citenamefont {F\`eve}}]{bocquillon:2013}%
  \BibitemOpen
  \bibfield  {author} {\bibinfo {author} {\bibfnamefont {E.}~\bibnamefont
  {Bocquillon}}, \bibinfo {author} {\bibfnamefont {V.}~\bibnamefont {Freulon}},
  \bibinfo {author} {\bibfnamefont {J.-M.}\ \bibnamefont {Berroir}}, \bibinfo
  {author} {\bibfnamefont {P.}~\bibnamefont {Degiovanni}}, \bibinfo {author}
  {\bibfnamefont {B.}~\bibnamefont {Pla\c{c}ais}}, \bibinfo {author}
  {\bibfnamefont {A.}~\bibnamefont {Cavanna}}, \bibinfo {author} {\bibfnamefont
  {Y.}~\bibnamefont {Jin}}, \ and\ \bibinfo {author} {\bibfnamefont
  {G.}~\bibnamefont {F\`eve}},\ }\bibfield  {title} {\enquote {\bibinfo {title}
  {Coherence and indistinguishability of single electrons emitted by
  independent sources},}\ }\href
  {http://www.sciencemag.org/content/339/6123/1054.abstract} {\bibfield
  {journal} {\bibinfo  {journal} {Science}\ }\textbf {\bibinfo {volume}
  {339}},\ \bibinfo {pages} {1054} (\bibinfo {year} {2013})}\BibitemShut
  {NoStop}%
\bibitem [{\citenamefont {Moskalets}(2012)}]{moskalets:book}%
  \BibitemOpen
  \bibfield  {author} {\bibinfo {author} {\bibfnamefont {M.~V.}\ \bibnamefont
  {Moskalets}},\ }\href@noop {} {\emph {\bibinfo {title} {Scattering matrix
  approach to non-stationary quantum transport}}}\ (\bibinfo  {publisher}
  {Imperial {C}ollege {P}ress},\ \bibinfo {year} {2012})\BibitemShut {NoStop}%
\bibitem [{\citenamefont {Dubois}\ \emph
  {et~al.}(2013{\natexlab{b}})\citenamefont {Dubois}, \citenamefont {Jullien},
  \citenamefont {Grenier}, \citenamefont {Degiovanni}, \citenamefont
  {Roulleau},\ and\ \citenamefont {Glattli}}]{dubois:2013prb}%
  \BibitemOpen
  \bibfield  {author} {\bibinfo {author} {\bibfnamefont {J.}~\bibnamefont
  {Dubois}}, \bibinfo {author} {\bibfnamefont {T.}~\bibnamefont {Jullien}},
  \bibinfo {author} {\bibfnamefont {C.}~\bibnamefont {Grenier}}, \bibinfo
  {author} {\bibfnamefont {P.}~\bibnamefont {Degiovanni}}, \bibinfo {author}
  {\bibfnamefont {P.}~\bibnamefont {Roulleau}}, \ and\ \bibinfo {author}
  {\bibfnamefont {D.~C.}\ \bibnamefont {Glattli}},\ }\bibfield  {title}
  {\enquote {\bibinfo {title} {Integer and fractional charge lorentzian voltage
  pulses analyzed in the framework of photon-assisted shot noise},}\ }\href
  {\doibase 10.1103/PhysRevB.88.085301} {\bibfield  {journal} {\bibinfo
  {journal} {Phys. Rev. B}\ }\textbf {\bibinfo {volume} {88}},\ \bibinfo
  {pages} {085301} (\bibinfo {year} {2013}{\natexlab{b}})}\BibitemShut
  {NoStop}%
\end{thebibliography}
\end{document}